\documentclass[a4paper,fleqn]{cas-sc}

\usepackage[authoryear,longnamesfirst]{natbib}
\usepackage{lineno}
\usepackage{subfigure}
\usepackage{float}
\restylefloat{figure}
\usepackage{amsmath}
\usepackage{bm}
\newtheorem{theorem}{Theorem}[section]
\usepackage[title]{appendix}
\usepackage{url}

\def\tsc#1{\csdef{#1}{\textsc{\lowercase{#1}}\xspace}}
\tsc{WGM}
\tsc{QE}

\begin{document}
\let\WriteBookmarks\relax
\def\floatpagepagefraction{1}
\def\textpagefraction{.001}
\let\printorcid\relax    
\shorttitle{Convex-PIML}   

\shortauthors{L. YI, S. YANG, Y. CUI, Z. LAI}  

\title [mode = title]{Transforming physics-informed machine learning to convex optimization}  

\author[1]{Letian YI}

\affiliation[1]{organization={Internet of Things Thrust},
            addressline={The Hong Kong University of Science and Technology(Gunagzhou)}, 
            city={Guangzhou},
            postcode={511453}, 
            state={Guangdong Province},
            country={China}}

\author[1]{Siyuan YANG}

\author[1,2]{Ying CUI}

\affiliation[2]{organization={Department of Electronic and Computer Engineering},
            addressline={The Hong Kong University of Science and Technology},
            city={Clear Water Bay},
            state={Hong Kong SAR},
            country={China}}
            
\author[1,3]{Zhilu LAI}
\cormark[1]

\affiliation[3]{organization={Department of Civil and Environmental Engineering},
            addressline={The Hong Kong University of Science and Technology},
            city={Clear Water Bay},
            state={Hong Kong SAR},
            country={China}}
            
\cortext[1]{Corresponding author. E-mail addresses: zhilulai@ust.hk}


\begin{abstract}
Physics-Informed Machine Learning (PIML) offers a powerful paradigm of integrating data with physical laws to address important scientific problems, such as parameter estimation, inferring hidden physics, equation discovery, and state prediction, etc. However, PIML still faces many serious optimization challenges that significantly restrict its applications. In this study, we transform PIML to convex optimization to overcome all these limitations, referred to as \textbf{Convex-PIML}. The linear combination of B-splines is utilized to approximate the data, promoting the convexity of the loss function. By replacing the non-convex components of the loss function with convex approximations, the problem is further converted into a sequence of successively refined approximated convex optimization problems. This conversion allows the use of well-established convex optimization algorithms, obtaining solutions effectively and efficiently. Furthermore, an adaptive knot optimization method is introduced to mitigate the spectral bias issue of PIML, further improving the performance. The proposed fully adaptive framework by combining the adaptive knot optimization and BSCA is tested in scenarios with distinct types of physical prior. The results indicate that optimization problems are effectively solved in these scenarios, highlighting the potential of the framework for broad applications.
\let\thefootnote\relax\footnotetext{GitHub: \url{https://github.com/YILotte/Convex-PIML}}
\let\thefootnote\relax\footnotetext{The manuscirpt was submitted to an international journal for peer review.}
\end{abstract}



\maketitle
\section{Introduction}\label{sec:introduction}
Physics-Informed Machine Learning (PIML) has recently garnered significant research interest \cite{tang2022physics, raabe2023accelerating, lai2021structural, kontolati2024learning, zhang2023mixed, vinuesa2022enhancing}, as a robust tool for integrating data with physical models. Among the methods for embedding physical laws in machine learning, soft constraints have become prominent due to their flexibility in incorporating diverse physics priors \cite{karniadakis2021physics}. A prime example of this approach is physics-informed neural networks (PINNs) \cite{raissi2019physics}, which have been extensively explored in various fields \cite{xu2023transfer, chen2023learning, xie2024physics, chen2021physics}.

\begin{figure}[H]
\centering 
\subfigure[The way of soft constraints to endow machine learning with physics.]{
\label{fig:soft constraint}
\includegraphics[width=0.8\textwidth]{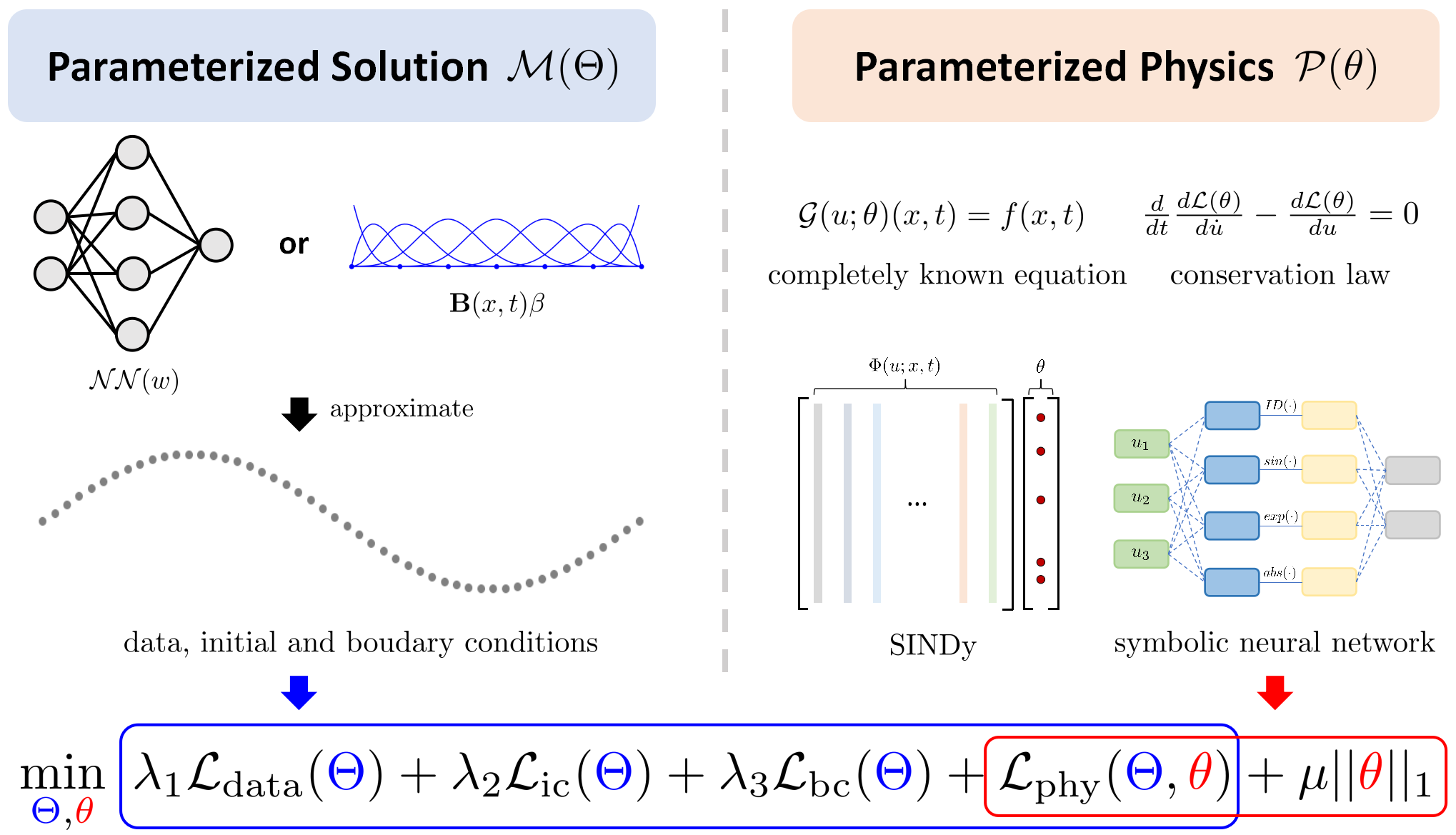}}
\subfigure[The optimization challenges of soft-constrained physics-informed machine learning.]{
\label{fig:optimization challenges}
\includegraphics[width=0.8\textwidth]{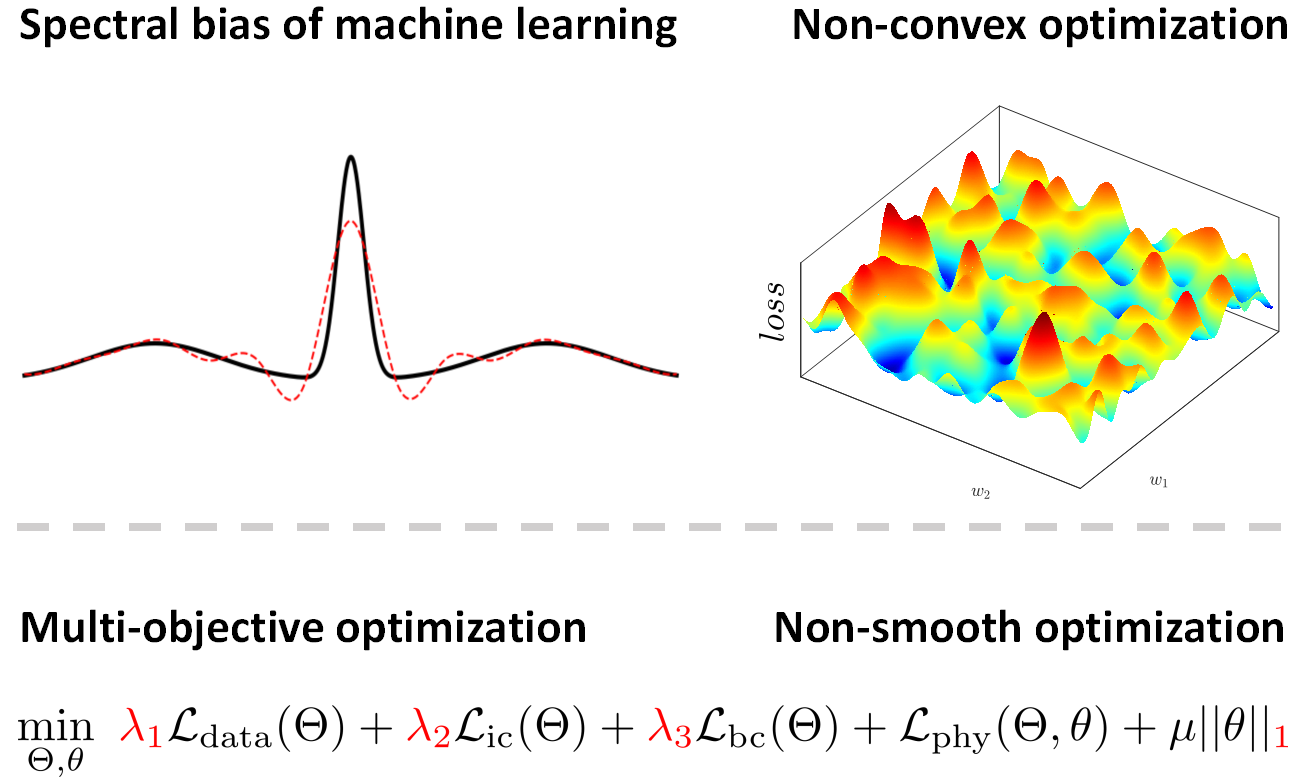}}
\caption{The soft-constrained physics-informed machine learning and its optimization challenges.}
\end{figure}

As illustrated in Fig.\ref{fig:soft constraint}, soft-constrained PIML relies on two key components: learnable functions used to fit the data, and underlying physics equations governing the data. Soft-constrained PIML is usually formulated as an optimization problem to minimize a loss function, including data fitting loss, initial condition loss, boundary condition loss, and physics loss. In equation discovery problems, an $\ell_1$ loss is added to promote sparsity in the discovered equation. The effectiveness of PIML hinges on the efficiency of solving this optimization problem. However, as presented in Fig.\ref{fig:optimization challenges}, significant challenges remain, such as spectral bias, non-convex optimization, multi-objective optimization, and non-smooth optimization, especially when dealing with nonlinear physics models and multi-scale data. This still makes applying PIML to solving scientific problems in various fields tricky.

The nonlinearity of machine learning and physics models often renders the PIML optimization a non-convex optimization problem. Non-convex optimization is notoriously challenging due to issues such as local minima, saddle points, and sensitivity to initialization \cite{jin2021nonconvex,dauphin2014identifying, jain2017non}. Some stochastic optimization algorithms, such as Stochastic Gradient Descent (SGD) and Adaptive Moment Estimation (Adam), incorporate noise and momentum, can mitigate these challenges to a certain extent by helping to escape local optima and saddle points \cite{ge2015escaping, kingma2014adam, sutskever2013importance}. Although sometimes local minima are good enough, PIML requires higher accuracy and robustness to extract precise and unique physical knowledge. Consequently, non-convex optimization remains a significant limitation for PIML. Most research focuses on the family of gradient descent methods due to their computational efficiency and effectiveness \cite{xiang2022self,zhang2022gw,jagtap2020conservative}. On the other hand, the gradient descent method also introduces other challenges, including multi-objective optimization related to gradient competition among loss terms and non-smooth optimization due to $\ell_1$ loss.

Regarding the multi-objective optimization of PIML, Wang et al. \cite{wang2021understanding} highlighted that inappropriate trade-off parameters could lead to imbalanced back-propagated gradient magnitudes across different loss terms, resulting in unbalanced optimization. Specifically, the data fitting loss, initial condition loss, and boundary condition loss are hard to minimize since the gradient of physics loss always dominates during training. They proposed a learning rate annealing algorithm that adjusts trade-off parameters based on statistical indices of gradient vectors. These indices may not accurately reflect the gradient magnitude, potentially failing to balance the gradients. Alternative methods like Dynamic Weight Average (DWA) \cite{liu2019end} and SoftAdapt \cite{heydari2019softadapt} focus on the convergence speed of loss instead of gradients to adjust trade-off parameters. However, they overlook the impact of loss magnitudes. GradNorm \cite{chen2018gradnorm} makes trade-off parameters trainable by defining a loss function that considers the convergence speed of loss and the $\ell_2$ norm of gradients. However, it increases computational complexity and introduces additional hyperparameters. A simple and broadly applicable method for determining trade-off parameters in PIML optimization remains elusive.

To promote the sparsity and interpretability of discovered equations, PIML must address the challenge of non-smooth optimization associated with the $\ell_0$ loss. However, solving the optimization problem with $\ell_0$ loss is not pratical since it is NP-hard. Two approaches are used to tackle it. The first involves approximating the $\ell_0$ loss with the non-convex $\ell_p\ (0<p<1)$ loss \cite{kim2020integration, chen2023learning}, which often leads to convergence to local minima. The second approach employs the convex $\ell_1$ loss to approximate the $\ell_0$ loss. The $\ell_{1}$ loss is inherently non-smooth, posessing a new optimization challenge. To tackle this non-smooth optimization, some well-established algorithms have been developed to solve specific problems with $\ell_1$ loss. For example, the Fast Iterative Shrinkage-Thresholding Algorithm (FISTA) \cite{beck2009fast}, based on the soft thresholding operator, proximal operator, and Nesterov accelerated gradient, addresses non-smooth optimization efficiently for the least-squares problems with a $\ell_1$ loss.

The spectral bias, refers to the tendency of learnable functions to learn different frequency components at varying rates. This issue is inherent in the model itself rather than the optimization algorithm. However, it also ultimately limits the optimization accuracy. In PINNs, spectral bias poses challenges to broader applications due to the neural network's preference for low-frequency content \cite{rahaman2019spectral, ronen2019convergence}. Various methods have been proposed to address the spectral bias of PINNs, including domain decomposition into subdomains \cite{moseley2023finite}, embedding Fourier features \cite{wang2021eigenvector}, and using self-adaptive sampling or weighting techniques \cite{gao2023failure, tang2023pinns, wu2023comprehensive, mcclenny2023self}. The spectral bias in neural networks can be analyzed with the neural tangent kernel theory, which is based on the infinite-width limit of networks \cite{jacot2018neural, arora2019exact, lee2019wide}. Local function-based methods use basis functions with local support to approximate solutions of physics equations. Notable examples include the works of Ramsay et al. \cite{ramsay2007parameter}, Frasso et al. \cite{frasso2016parameter}, and Bhowmick et al. \cite{bhowmick2023data}, which focus on parameter estimation of differential equations. However, these studies do not address the spectral bias of basis functions. The local support of basis functions offers potential for adjustment to enhance local approximations. Established theories of error-bound analysis and the explicit characteristics of basis functions can theoretically guide these adjustments to mitigate spectral bias. Despite this, most research on the spectral bias of basis functions has focused on data fitting problem \cite{lan2024full, shi2023residuals, jiang2022scattered}, leaving the problem with physics constraints unexplored.

Due to the advantageous properties of convex optimization, transforming physics-informed machine learning into convex optimization holds considerable promise for mitigating all the challenges mentioned above. One of the most important advantages of convex optimization is that, in general, globally optimal solutions can be obtained easily, irrespective of initial points \cite{boyd2004convex}. To leverage this advantageous property and to alleviate the non-convex optimization issues, we can enhance the convexity of the loss function by utilizing the linear combination of basis functions to fit the data, leading to a convex data loss function. If the non-convex part of loss function (i.e., the physics loss) can be further approximated by carefully chosen convex functions, the original non-convex loss function can be transformed into the combination of some convex functions. In this case, there is no need to use dynamic gradient balancing techniques, since the balance between convex losses can be easily maintained by assigning a large constant weight to the loss function difficult to optimize (we will demonstrate it in Methods). Furthermore, transforming PIML into convex optimization problems also allows us to utilize well-established convex optimization algorithms with strong convergence properties to effectively find solutions in polynomial time.

In this paper, we propose a comprehensive framework based on convex optimization, referred to as Convex-PIML, to address the aforementioned issues by (1) employing block coordinate descent to separately update $\Theta$ (parameterizing solutions) and $\theta$ (parameterizing informed physics, as shown in Fig.\ref{fig:soft constraint}), thereby simplifying the problem; (2) utilizing a linear combination of B-splines to fit the data, which makes the loss function weakly non-convex, significantly improving the loss landscapes (illustrated in Fig.\ref{fig:loss landscape bspline}); and (3) approximating the non-convex component of the loss function—specifically, the physics loss—with a well-designed convex function, thereby transforming the problem into a sequence of successively refined convex optimization problems. This enables the efficient and effective application of convex optimization; (4) proposing an adaptive knot optimization method to solve the spectral bias. The efficiency of Convex-PIML are demonstrated across various scenarios with differing levels of physics priors, including known equations, equations represented by sparse identification of nonlinear dynamics (SINDy) \cite{brunton2016discovering}, and conservation laws represented by the Euler-Lagrange equation.

\section{Methods}\label{sec:methods}
\subsection{Formulation of the optimization problem}\label{sec:formulation}
The governing equation in a general form is given as:
\begin{equation}
\label{eq:governing equation}
\mathcal{F}(u; \theta)(x,t)=0,
\end{equation}
where $u(x,t)$ denotes the solution to this equation; $x=[x_1,x_2,...,x_n]^{\text{T}}$ denotes the spatial dimensions, and $t$ represents the temporal dimension; the equation is considered to be parameterized by $\theta=[\theta_1,...,\theta_m]^{\text{T}}$; and $\mathcal{F}(\cdot)$ is a continuous operator acting on $u(x,t)$.

Given measurements $\mathbf{y} = [y_1,y_2,...,y_{N_d}]^{\text{T}}$ of $u(x,t)$ governed by Eq.\eqref{eq:governing equation}, it aims to use the proposed scheme to effectively and robustly achieve different objectives corresponding to different physics priors: (1) given known equation, to estimate parameters, and further induce hidden physics (quantities such as $\frac{\partial u}{\partial t}, \frac{\partial^2 u}{\partial x_1 \partial x_2}, \frac{\partial^2 u}{\partial x_1 \partial t}$, etc.); (2) given partially known or unknown equations, to discover unknown parts from data; (3) given the general conservation law, to train a surrogate model of this law and then predict system state. All above objectives can be unified by solving the following optimization problem:
\begin{equation}
\label{eq:contrained optimization}
\begin{aligned}
    \mathop{\min}\limits_{u,\theta}\sum_{i=1}^{N_d}&\big(y_i-u(x^d_i,t^d_i)\big)^2,\\
      \text{s.t.}\    \mathcal{F}(u;\theta)(x,t)&=0, x\in\Omega,\\
  u(x,0)&=\phi(x),\\
u(x,t)&=\psi(x,t),x\in\partial \Omega,
\end{aligned}
\end{equation}
where $\Omega$ and $\partial \Omega$ are the domain and boundary of the governing equation, respectively; $x^d=[x^d_1,x^d_2,...,x^d_{N_d}]^{\text{T}}$ and $t^d=[t^d_1,t^d_2,...,t^d_{N_d}]^{\text{T}}$ are the coordinate vectors of measured data. The closed-form $u(x,t)$ is commonly difficult to obtain as either the governing equation or the constrained optimization is hard to solve. Therefore, one can use a family of functions $\mathcal{M}(\Theta)$ parameterized by $\Theta$  to obtain an approximation $\hat{u}(x,t)=\mathcal{M}(x,t;\Theta^*)\approx u(x,t)$, where $\Theta^*$ denotes the optimal parameters. By applying the operator $\mathcal{F}(\cdot)$ to the approximation $\hat{u}(x,t)$, the governing equation is turned into $\mathcal{F}(\hat{u};\Theta^*,\theta^*)(x,t)=0, x\in\Omega$, i.e., $\mathcal{F}(\cdot)$ is parameterized by both $\Theta$ and $\theta$.

To determine the $\Theta^*$ and $\theta^*$, the constrained optimization is transformed into an unconstrained optimization, which is a weighted sum of all loss terms:
\begin{equation}\label{eq:unconstrained optimization}
\mathop{\min}\limits_{\Theta,\theta}\ \lambda_1\mathcal{L}_\text{data}(\Theta)+\lambda_2\mathcal{L}_\text{ic}(\Theta)+\lambda_3\mathcal{L}_\text{bc}(\Theta)+\mathcal{L}_\text{phy}(\Theta,\theta)+\mu||\theta||_1,
\end{equation}
where $\lambda_1,\lambda_2,\lambda_3$ are trade-off parameters that balance the contributions of all loss terms; $\mu||\theta||_1$ denotes the weighted $\ell_1$ loss to promote the sparsity of discovered equation ($\mu = 0$ when no sparsity of $\theta$ is introduced); $\mathcal{L}_\text{data}$ denotes the data fitting loss $\mathcal{L}_\text{data}(\Theta)=\frac{1}{2N_d}||\mathbf{y}-\mathcal{M}(x^d,t^d;\Theta)||_2^2$. Similarly, the initial condition loss $\mathcal{L}_\text{ic}(\Theta)=\frac{1}{2N_{\text{ic}}}||\phi-\mathcal{M}(x^\text{ic},t^\text{ic};\Theta)||_2^2$ and boundary condition loss $\mathcal{L}_\text{bc}(\Theta)=\frac{1}{2N_{bc}}||\psi-\mathcal{M}(x^\text{bc},t^\text{bc};\Theta)||_2^2$ have the same form. $\mathcal{L}_\text{phy}(\Theta,\theta)$ denotes the physics loss $\iint \big(\mathcal{F}(\hat{u};\Theta,\theta)(x,t) \big)^2dxdt$.

The integral form of this physics loss is however not easy to optimize, and one can define the following summation form as an approximation $\mathcal{L}_\text{phy}(\Theta,\theta)=\frac{1}{2N_c}||\mathcal{F}(\hat{u};\Theta,\theta)(x^c,t^c)||^2_2$, where $x^c=[x^c_1,x^c_2,...,x^c_{N_c}]$ and $t^c=[t^c_1,t^c_2,...,t^c_{N_c}]$ are the coordinate vectors of \textit{collocation points}, at which the approximated solution ($\hat{u}^c = \hat{u}(x^c, t^c)$) is required to satisfy the governing equation. Note that the coordinates of collocation points ($x_i^c, t_i^c$) are not necessarily the same as the coordinates of data points ($x_i^d, t_i^d$).
Letting $g(\Theta)=\lambda_1\mathcal{L}_\text{data}(\Theta)+\lambda_2\mathcal{L}_\text{ic}(\Theta)+\lambda_3\mathcal{L}_\text{bc}(\Theta)$ and $h(\Theta,\theta)=\mathcal{L}_\text{phy}(\Theta,\theta)$, the optimization problem is expressed as:
\begin{equation}\label{eq:unconstrained optimization BSCA}
\mathop{\min}\limits_{\Theta,\theta}\ g(\Theta)+h(\Theta,\theta)+\mu||\theta||_1.
\end{equation}
\subsection{B-spline basis functions}\label{sec:b-splines}
Due to the strong nonlinearity of neural networks $\mathcal{NN}(w)$, the loss landscape of PINNs (here $\Theta=w$ and $\theta$ is fixed as the value at the $(k-1)^{\text{th}}$ iteration of gradient descent) is strongly non-convex, as illustrated in Fig.\ref{fig:loss landscape nn}.

To prompt the convexity of the loss landscape, the linear combination of basis function $\mathbf{B}(x,t)\beta$ is chosen as learnable function (i.e., $\Theta=\beta$) to approximate $\mathcal{M}(x,t;\Theta)$, where $\mathbf{B}(x,t)=[b_1(x,t),...,b_i(x,t),...,b_N(x,t)]$ comprises a series of basis functions and $\beta=[\beta_1,...,\beta_i,...,\beta_N]^{\text{T}}\in \mathbb{R}^{N}$ is a vector consists of respective weight coefficients for each basis function. Then $g(\beta)$ is convex. The initial and boundary conditions uniquely determine the solution to differential equation. Besides, data must be well fitted. To satisfy these two requirements, trade-off parameters should be large to minimize $g(\beta)$ in PIML. In this case of dominated $g(\beta)$, the loss function $g(\beta)+h(\beta,\theta)$ will be a weakly non-convex function. As shown in Fig.\ref{fig:loss landscape bspline}, the loss landscape is much more smoother and more convex than PINNs. 

As for the specific basis functions, we choose B-splines generated based on predefined knots. B-splines have the advantages of smoothness, flexibility, and stability in numerical calculations \cite {piegl2012nurbs}. More importantly, the features of B-splines can be adjusted by moving and refining the knots, as shown in Fig.\ref{fig:adjusting knots}. As a rule of thumb, spiky B-splines are required to capture spiky data features. We design a knot optimization algorithm (including knot movement and knot refinement) to match B-spline features with the features of functions to be approximated. It can significantly improve the accuracy of the approximation, as shown in Fig.\ref{fig:knot optimization}. 

Moreover, unlike PINNs using auto-differentiation to obtain derivatives, the accurate (partial) derivative of B-splines can be explicitly derived (such as $\frac{\partial \mathbf{B}(x,t)}{\partial x}$, or $\frac{\partial \mathbf{B}(x,t)}{\partial t}$, etc.). It greatly reduces computational demands and improves convergence \cite{kaewnuratchadasorn2024physics}. To accommodate high-dimensional problems, B-splines with high-dimension can be generated by tensor products of one-dimensional B-splines \cite{bhowmick2023data}.
\subsection{Block successive convex approximation (BSCA)}\label{sec:bsca}
Considering the structure of the loss function in Eq.\eqref{eq:unconstrained optimization}, the block coordinate descent method is adopted. Specifically, this optimization can be greatly simplified by dividing the $\beta$ and $\theta$ into two blocks to be optimized alternately. The sub-problem of obtaining $\beta^{(k)}$ (the $k^{\text{th}}$ iteration of updating $\beta$) is given as:
\begin{equation}\label{eq:bsca beta}
    \beta^{(k)}=\mathop{\arg \min\limits_{\beta}} \ g(\beta)+h(\beta,\theta^{(k-1)}).
\end{equation} 
After solving the sub-problem to obtain $\beta^{(k)}$, the sub-problem of obtaining $\theta^{(k)}$ should be solved to update $\theta$ as well:
\begin{equation}\label{eq:update theta}
    \theta^{(k)} =\mathop{\arg \min\limits_{\theta}} \ h(\beta^{(k)},\theta)+\mu||\theta||_1.
\end{equation}
\subsubsection*{When $h(\beta,\theta)$ is convex}
If the physics loss $h(\beta,\theta)$ is convex with respect to both $\beta$ and $\theta$, the sub-problem can be directly solved by convex optimization algorithms. For example, Eq.\eqref{eq:bsca beta} can be solved by the Alternative Direction Method of Multipliers (ADMM) \cite{gabay1976dual}; Eq.\eqref{eq:update theta} can be solved by Fast Iterative Shrinkage-Thresholding Algorithm (FISTA) \cite{beck2009fast}. 
\subsubsection*{When $h(\beta,\theta)$ is non-convex}
When $h(\beta,\theta)$ is non-convex with respect to $\theta$ or $\beta$, a well-designed convex function is adopted to approximate $h(\beta,\theta)$ in each iteration of solving sub-problems. Specifically, the quadratic approximation~\cite{yang2019inexact} is used in our study (using the process of updating $\beta$ as an example):
\begin{equation}\label{eq:quadratic appro}
    \tilde{h}(\beta,\theta^{(k-1)})=(\beta-\beta^{(k-1)})^{\text{T}}\nabla_{\beta}{h}(\beta^{(k-1)},\theta^{(k-1)})+\frac{c}{2}||\beta-\beta^{(k-1)}||_2^2,
\end{equation}
where $\nabla_{\beta}{h}(\beta^{(k-1)},\theta^{(k-1)})$ is the gradient of $h(\beta,\theta^{(k-1)})$ evaluated at $\beta^{(k-1)}$; $c$ is a positive scalar. Then the approximated solution of Eq.\eqref{eq:bsca beta} can be obtained by solving the following convex problem:
\begin{equation}\label{eq:approx beta}
    \tilde{\beta}^{(k)}=\mathop{\arg \min\limits_{\beta}} \ g(\beta)+\tilde{h}(\beta,\theta^{(k-1)}),
\end{equation}
where $\tilde{\beta}^{(k)}$ is the approximation of ${\beta}^{(k)}$. Finally, $\beta^{(k)}$ can be obtained, where $\gamma^{(k)}$ is a step size at the $k^{\text{th}}$ step:
\begin{equation}\label{eq:update beta}
    \beta^{(k)}=\beta^{(k-1)}+ \gamma^{(k)}(\tilde{\beta}^{(k)}-\beta^{(k-1)}).
\end{equation}

The updating process of $\theta$ is the same as $\beta$. Note the non-smooth sub-problem of updating $\theta$ can be directly solved by FISTA \cite{beck2009fast} since it has been converted to a convex problem. This updating scheme is known as Block Successive Convex Approximation (BSCA) \cite{yang2019block}.

Then we will show how BSCA solves the optimization challenges for PIML. Note that $g(\beta)=\frac{\lambda}{2N_d}||\mathbf{y}-\mathbf{B}^d\beta||_2^2$ for simplicity, where $\mathbf{B}^d=\mathbf{B}(x^d,t^d)\in \mathbb{R}^{N_d\times N}$. This means that the information on initial and boundary conditions is perfectly given in the data. Thus, the initial and boundary condition losses can be combined into data fitting loss. The gradient of $g(\beta)$ is $\frac{\lambda}{N_d}(-\mathbf{B}^d\mathbf{y}+(\mathbf{B}^d)^{\text{T}}\mathbf{B}^d\beta)$ and the gradient of $\tilde{h}(\beta,\theta^{(k-1)})$ is $\nabla_{\beta}{h}(\beta^{(k-1)},\theta^{(k-1)})+c(\beta-\beta^{(k-1)})$. Then the \textit{analytical solution} of $\tilde{\beta}^{(k)}$ is obtained by setting the derivative to zero to solve the least squares problem in Eq.\eqref{eq:approx beta}: 
\begin{equation}\label{eq:tilde beta bsca}
      \tilde{\beta}^{(k)}=\left(\frac{\lambda}{N_d}(\mathbf{B}^d)^{\text{T}}\mathbf{B}^d+c\mathbf{I}\right)^{-1}\left(\frac{\lambda}{N_d}(\mathbf{B}^d)^{\text{T}}\mathbf{y}+c\beta^{(k-1)}-\nabla_{\beta}{h}(\beta^{(k-1)},\theta^{(k-1)})\right),
\end{equation}
where $\mathbf{I}\in\mathbb{R}^{N\times N}$ is the unit diagonal matrix. For simplicity, we assume $\gamma^{(k)}=1$, thus ${\beta}^{(k)}=\tilde{\beta}^{(k)}$. 

When $\lambda=0$ (meaning that the loss related to data is set to be zero, and one only optimizes the loss related to physics), Eq.\eqref{eq:tilde beta bsca} is equivalent to using gradient descent with learning rate of $\frac1c$ to update $\beta$, i.e., 
\begin{equation}\label{eq:tilde beta gd}
   {\beta}^{(k)}=\beta^{(k-1)}-\frac{1}{c}\nabla_\beta h(\beta^{(k-1)},\theta^{(k-1)}). 
\end{equation}

\begin{figure}[H]
\centering 
\subfigure[The illustration of strongly non-convex loss landscapes using neural networks.]{
\label{fig:loss landscape nn}
\includegraphics[width=0.7\textwidth]{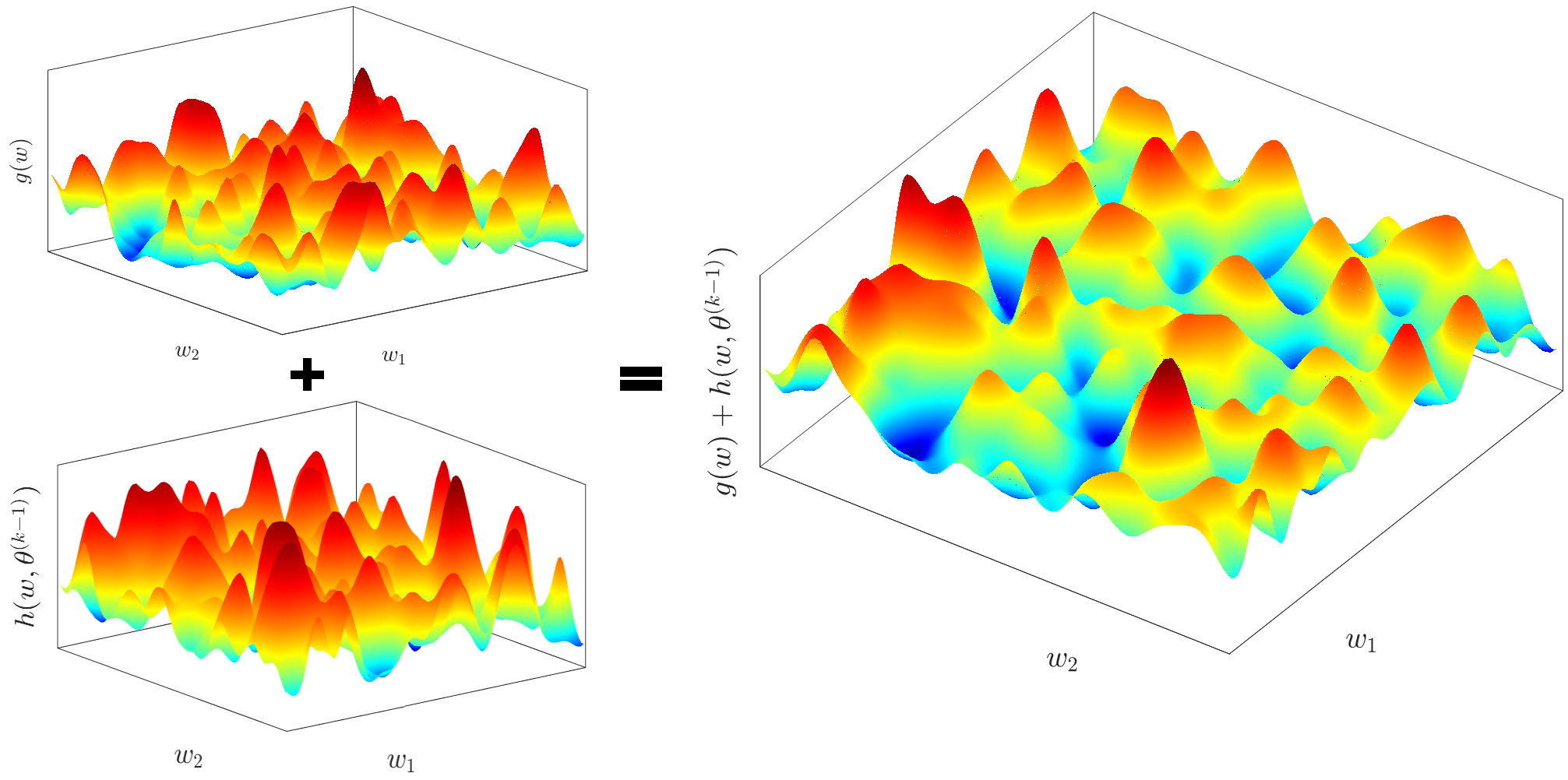}}
\subfigure[The weakly non-convex loss landscape with the linear combination of basis functions.]{
\label{fig:loss landscape bspline}
\includegraphics[width=0.7\textwidth]{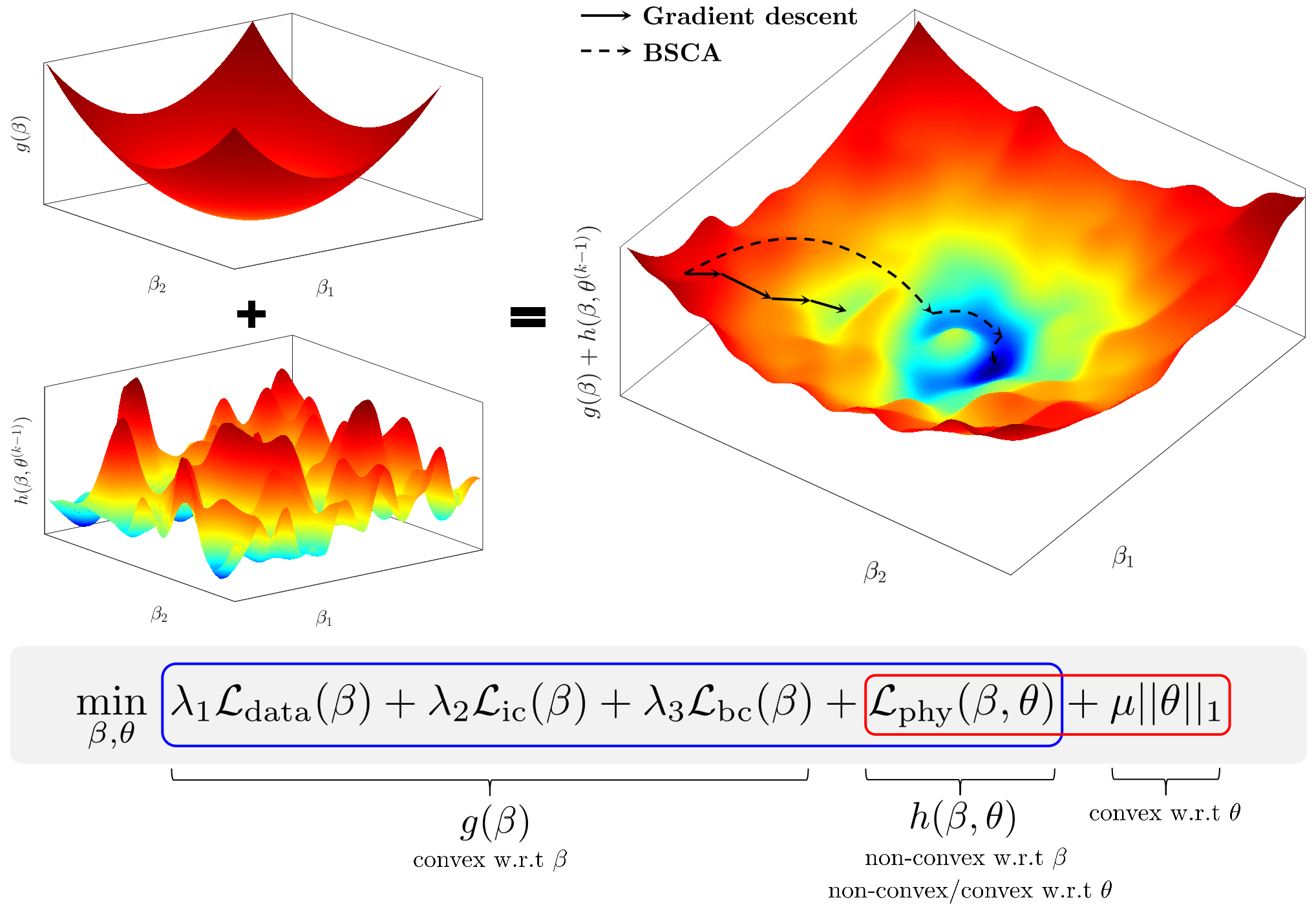}}
\subfigure[The update scheme of BSCA, where each subproblem is a convex optimization by using quadratic approximation.]{
\label{fig:BSCA}
\includegraphics[width=0.7\textwidth]{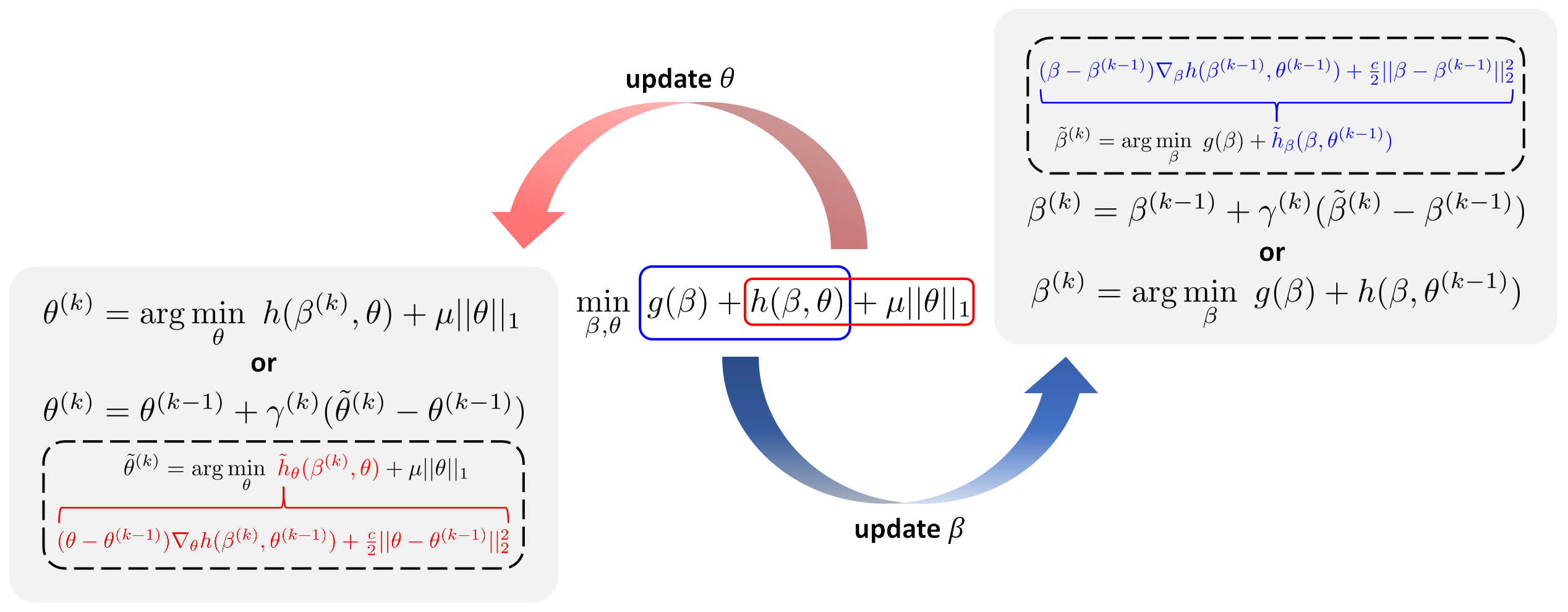}}
\caption{The comparison of loss landscapes with the neural network and linear combination of basis functions and the comparison of updating method of gradient descent and BSCA.}
\end{figure}

It means that in this setting  BSCA uses a gradient descent-like approach to minimize the non-convex loss $h(\beta,\theta)$. As for the convex loss $g(\beta)$, if the trade-off parameter $\lambda$ is large enough to make $g(\beta)$ dominate, the solution is close to the solution of $\mathop{\min\limits_{\beta}}\frac{\lambda}{2N_d}||\mathbf{y}-\mathbf{B}^d\beta||_2^2$, i.e., Eq.\eqref{eq:tilde beta bsca} reduces to $\tilde{\beta}^{(k)}\approx\big((\mathbf{B}^d)^{\text{T}}\mathbf{B}^d\big)^{-1}\big(\mathbf{B}^d\big)^{\text{T}}\mathbf{y}$. This can be viewed as a \textit{jumping} update method, jumping from arbitrary initial point to the trough of loss landscape, as shown in Fig.\ref{fig:loss landscape bspline}.

It should be noted that during the jumping update, BSCA still manages to optimize the non-convex term due to the regularization of $\tilde{h}(\beta,\theta^{(k-1)})$ in Eq.\eqref{eq:approx beta}. This jumping update method enables BSCA to continuously jump out of local minimums. In contrast, the gradient descent method is more like a walking update, as shown in Fig.\ref{fig:loss landscape bspline}. Thus gradient descent method is very sensitive to initial values and tends to stuck in a bad local minima. Besides, in BSCA the trade-off parameters are much easier to determine -- tuned to be sufficiently large. 

To make the optimization converge as quickly as possible, the exact line search is used to determine the optimal step size in our study. It is based on the local geometry of the loss landscape \cite{yang2019inexact}:
\begin{equation}\label{eq:exact line search}
        \gamma^{(k)}=\mathop{\arg \min}\limits_{0\leq\gamma\leq1}\ g(\beta^{(k-1)}+\gamma(\tilde{\beta}^{(k)}-\beta^{(k-1)}))+h(\beta^{(k-1)}+\gamma(\tilde{\beta}^{(k)}-\beta^{(k-1)}),\theta^{(k-1)}).
\end{equation}

The traditional exact line search usually suffers from a high complexity when dealing with non-smooth optimization problems, as the optimization problem in Eq.\eqref{eq:exact line search} is non-differentiable. As demonstrated in \cite{yang2017unified}, performing an exact line search on the following differentiable function can also produce an effective step size:
\begin{equation}\label{eq:modified exact line search}
\begin{aligned}
        \gamma^{(k)}&=\mathop{\arg \min}\limits_{0\leq\gamma\leq1}\ \gamma(g(\tilde{\beta}^{(k)})-g(\beta^{(k-1)}))+h(\beta^{(k-1)}+\gamma(\tilde{\beta}^{(k)}-\beta^{(k-1)}),\theta^{(k-1)})\\
        &=\mathop{\arg \min}\limits_{0\leq\gamma\leq1}\ s(\gamma).
\end{aligned}
\end{equation}

As $g(\beta)$ is convex, the objective function in \eqref{eq:modified exact line search} is an upper bound of the objective function in Eq.\eqref{eq:exact line search}: $g(\beta^{(k-1)}+\gamma(\tilde{\beta}^{(k)}-\beta^{(k-1)}))\leq(1-\gamma)g(\beta^{(k-1)})+\gamma g(\tilde{\beta}^{(k)})=\gamma(g(\tilde{\beta}^{(k)})-g(\beta^{(k-1)}))+g(\beta^{(k-1)})$, where $g(\beta^{(k-1)})$ is a constant. Eq.\eqref{eq:modified exact line search} is to find the solution in $[0,1]$ such that $\frac {ds(\gamma)}{d\gamma} = 0$, which is not difficult to solve using numerical methods. However, it may reduce computational efficiency. A much simpler but effective step size rule called diminishing step size \cite{scutari2013decomposition} can be alternatively used:
\begin{equation}\label{eq:step_size}
    \gamma^{(k)} \in (0,1],\ \lim_{k\rightarrow+\infty}\gamma^{(k)}=0, \ \sum_{k}\gamma^{(k)}=+\infty.
\end{equation}

Many diminishing step size rules in the literature \cite{bertsekas2003parallel} satisfy this rule. For instance, the following rule is found to be very effective in our experiments \cite{di2016next}:
\begin{equation}\label{eq:diminishing step size}
    \gamma^{(k)}=\gamma^{(k-1)}(1-\epsilon\gamma^{(k-1)}),\ \text{with} \ \gamma^{(0)}=1,
\end{equation}
where $\epsilon \in (0,1)$ is a user-defined constant.
\subsection{Adaptive knot optimization}\label{sec:knot optimization}
As mentioned in Section \ref{sec:b-splines}, spiky-shape (high-curvature) B-splines should be used to capture high-frequency features during approximation. Guided by this rule of thumb, we develop an \textit{adaptive knot optimization} algorithm to improve the approximation accuracy by adaptively adding and moving knots, since knots are essential in defining the curvature and continuity of the B-spline curve/surface, as shown in Fig. \ref{fig:knot optimization}. Specifically, smaller knot intervals lead to higher curvature in the B-spline basis functions, resulting in sharper and more localized shapes.

In contrast to purely data-driven problems, the loss function of PIML contains physics loss $\mathcal{L}_\text{phy}$. Therefore, we first analyze the fundamental optimization objective. The data $\mathbf{y}$ represents a noisy discretization of the true solution $u(x,t)$. By approximating the integral form with a summation form loss function, minimizing the data fitting loss can be interpreted as minimizing $||\hat{u}(x,t)-u(x,t)||_2^2$ -- the error of approximating $u(x,t)$ using basis functions. Minimizing the physics loss is equivalent to employing least-squares finite element methods to solve the strong form of PDEs \cite{monk1999least}. We propose an error estimate of least-squares finite element methods based on a posteriori error estimate developed by R.Verfürth \cite{verfurth1994posteriori,verfurth1996review} (further details can be found in Appendix \ref{sec:a psteriori}). It indicates that physics loss is an upper bound for $||\hat{u}(x,t)-u(x,t)||_2^2$. In this regard, minimizing the physics loss can also be viewed as minimizing $||\hat{u}(x,t)-u(x,t)||_2^2$. In conclusion, the optimization objective of PIML is to essentially minimize $||\hat{u}(x,t)-u(x,t)||_2^2$ as well. Building on this key insight, we can still apply the rule of thumb--spiky-shape B-splines should be used to capture high-frequency features--in our PIML optimization.

\subsubsection*{Knot refinement}
Our optimization starts with smooth B-splines defined on uniformly sparse knot distributions. The regions with high error in the data fitting distribution $e_d(x^d,t^d)$ and physics error distribution $e_p(x^c,t^c)$ typically indicate that the B-splines in these areas are too smooth to approximate $u(x,t)$. Therefore, the knot refinement adds new knots at the middle of knot intervals with higher cumulative data fitting and physics errors. Given a 2-D case with $p\times q$ knot mesh, the domain is divided into rectangular subdomains $\bm{\delta}_{ij}=[x_i,x_{i+1}]\times[t_j,t_{j+1}],\ i=0,...,p-1,\ j=0,...,q-1$. The cumulative data fitting error on the ${i}^{\text{th}}$ column in $x$ direction is defined as:
\begin{subequations}\label{eq19}
\begin{equation}\label{eq19a}
E_d^{x_i}=\sum_{(x^d,t^d)\in [x_i,x_{i+1}]\times[t_0,t_{q-1}]}e_d(x^d,t^d),\  0\leq i\leq p-1,
\end{equation}
and $E_d^{t_j}$ is the ${j}^{\text{th}}$ row in $t$ direction is defined as:
\begin{equation}\label{eq19b}
E_d^{t_j}=\sum_{(x^d,t^d)\in[x_0,x_{p-1}]\times[t_j,t_{j+1}]}e_d(x^d,t^d),\  0\leq j\leq q-1,
\end{equation}
\end{subequations}
The cumulative physics error has the similar form.

\subsubsection*{Knot movement}
The above refined knots may not result in B-splines whose curvature match the features of $u(x,t)$. Knot intervals should be increased in smooth regions of $u(x,t)$ to ensure smooth B-splines, and reduced in spiky regions to allow B-splines to better capture sharp features. This is equivalent to encouraging the cumulative curvature of $u(x,t)$ across all subdomains $\bm{\delta}_{ij}$ to be distributed as uniformly as possible. Therefore, the knot movement utilizes the curvature of $\hat{u}(x,t)$ ($u(x,t)$ is unkonwn) to optimize the position of knots. We inherit the objective function $D_r(x, t)$ proposed in \cite{zhang2016b} for dimension  $r$ (here $r=x$ or $t$), which reflects the variance of the cumulative curvature of $\hat{u}(x,t)$ in all subdomains $\bm{\delta}_{ij}$:
\begin{subequations}
    \begin{equation}
    \label{eq16}
        D_r(x, t)=\sum_{i=0}^{p-1}\sum_{j=0}^{q-1}(\int_{\bm{\delta}_{ij}}\kappa_r(x,t)dxdt-M_r)^2,
\end{equation}
\begin{equation}
\label{eq17}
    M_r=\frac{\sum_{i=0}^{p-1}\sum_{j=0}^{q-1}\int_{\bm{\delta}_{ij}}\kappa_r(x,t)dxdt}{pq},
\end{equation}
\end{subequations}
where $\kappa_r(x,t) =|\frac{\partial^2 \hat{u}}{\partial r^2}/(1+\frac{\partial \hat{u}}{\partial r})^{3/2}|$ is the curvature in the dimension $r$. $M_r$ is mean value of cumulative $\kappa_r(x,t)$. 

The partial derivatives of $D_r(x, t)$ with respect to knot $x_i$ and $t_j$ are computed as follows \cite{zhang2016b}, respectively:
\begin{subequations}\label{eq18}
\begin{equation}\label{eq18a}
    \frac{\partial D_x}{\partial x_i}=2\sum_{j=0}^{p-1}(\int_{\bm{\delta}_{i-1,j}}\kappa_x(x,t)dxdt-\int_{\bm{\delta}_{i,j}}\kappa_x(x,t)dxdt)\int_{t_{j}}^{t_{j+1}}\kappa_x(x_i,t)dt,
\end{equation}
\begin{equation}\label{eq18b}
    \frac{\partial D_t}{\partial t_j}=2\sum_{i=0}^{q-1}(\int_{\bm{\delta}_{i,j-1}}\kappa_t(x,t)dxdt-\int_{\bm{\delta}_{i,j}}\kappa_t(x,t)dxdt)\int_{x_{i}}^{x_{i+1}}\kappa_t(x,t_j)dx.
\end{equation}
\end{subequations}

\begin{figure}[H]
\centering 
\subfigure[The features of B-splines are controlled by the number and positions of knots.]{
\label{fig:adjusting knots}
\includegraphics[width=0.85\textwidth]{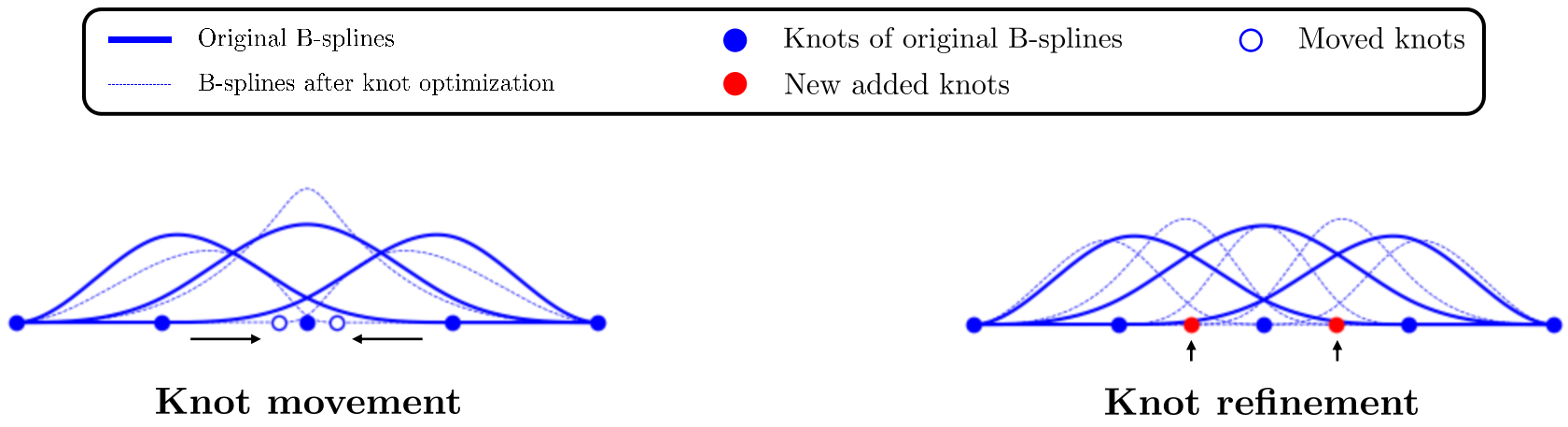}}
\subfigure[The knot optimization (knot movement and refinement) can improve the accuracy of approximation by matching the features of B-splines with true solution.]{
\label{fig:knot optimization}
\includegraphics[width=0.85\textwidth]{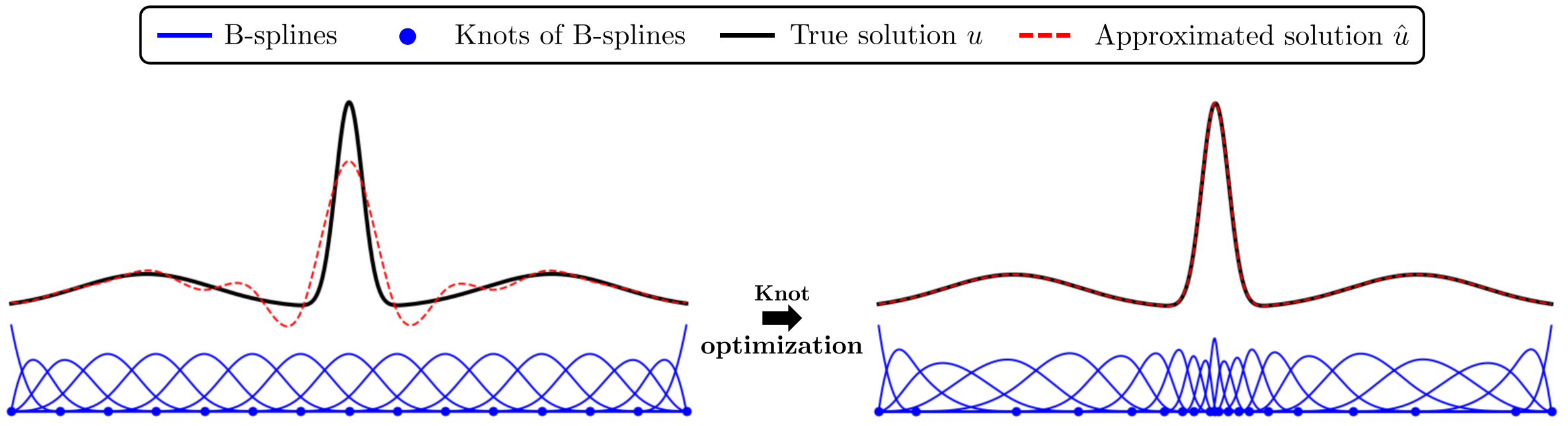}}
\subfigure[The proposed fully adaptive scheme where knots are dynamically adjusted during solving PIML optimization with BSCA.]{
\label{fig:framework}
\includegraphics[width=0.85\textwidth]{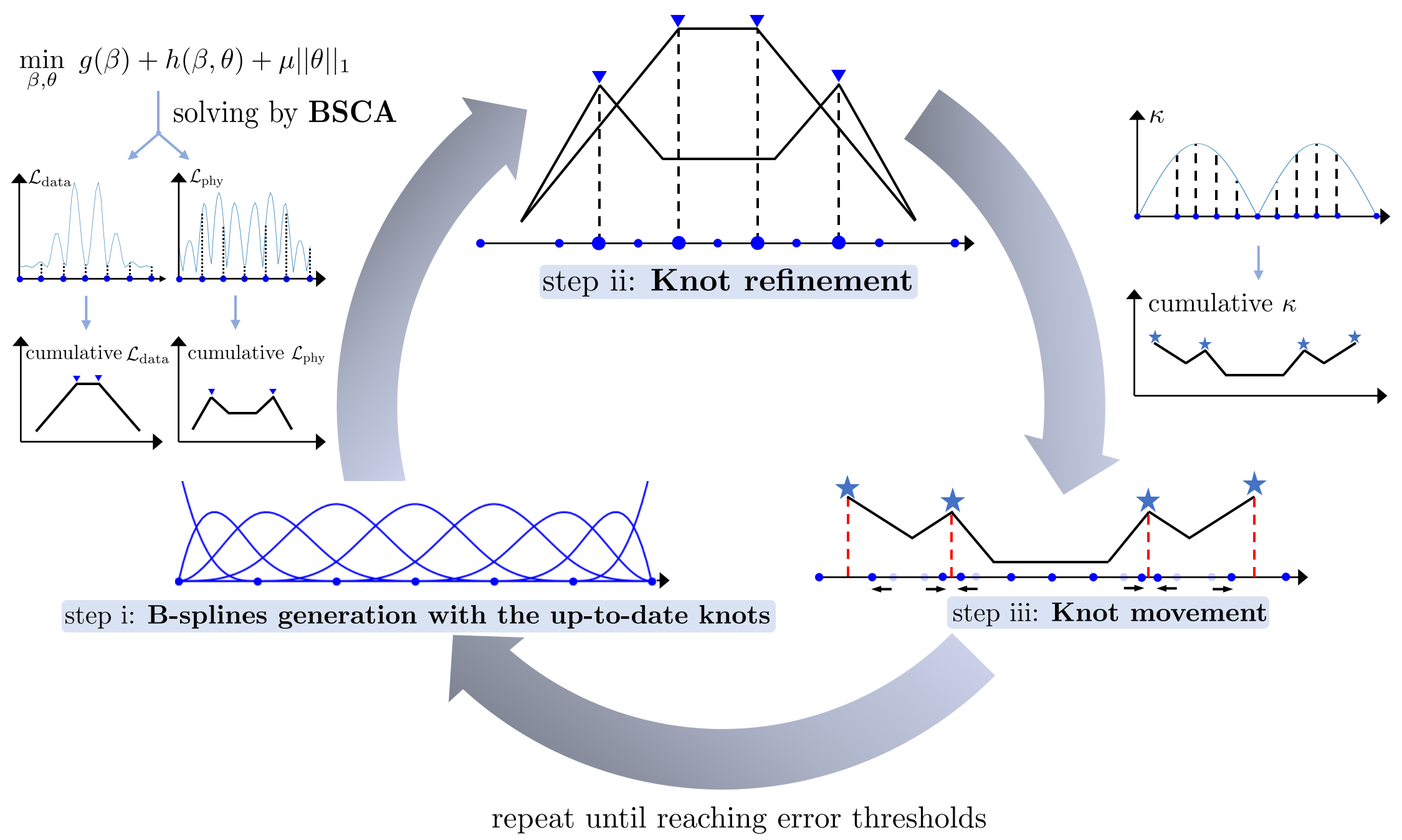}}
\caption{The knot optimization of B-splines and the proposed fully adaptive scheme.}
\label{fig:knots}
\end{figure}

Then we use gradient descent to optimize the position of knots. Since calculating the integral is time-consuming, in practice we use summation to approximate the integral. The high/low-frequency features of optimized B-splines agree with the true solution by knot refinement and movement, as illustrated in Fig.\ref{fig:knot optimization}.

We propose a fully adaptive scheme by combining the adaptive knot optimization and BSCA. We take a 1-D problem to illustrate the fully adaptive scheme in detail. Firstly, a series of B-spines are generated based on predefined sparse knots. Then the optimization problem is solved by BSCA to obtain error distributions and approximated solution $\hat{u}(t)$. New knots are inserted into knot intervals with greater cumulative data fitting error and cumulative physics error, as illustrated in Fig.\ref{fig:framework}. The positions of refined knots are further optimized in knot movement. Subsequently, a new series of B-splines are generated based on up-to-date knots. The above procedure is repeated until reaching predefined error threshold. Note that all components of this adaptive knot optimization are automated.

\section{Results}\label{sec:results}
The specific form of physics loss $h(\beta,\theta)$ in Eq.\eqref{eq:unconstrained optimization BSCA} varies given different physics priors. Physics priors can be categorized based on their degree of specificity. We evaluated the performance of the proposed method in scenarios with different physics priors to illustrate the broad applicability of Convex-PIML. The specifics of all following experiments can be found in the Appendix \ref{appendix:The experimental specifics}.
\subsection{Parameter estimation with known equations}
If the functional form of the governing equation is known as $\mathcal{G}(u; \theta)(x,t)=f(x,t)$, where $f(x,t)$ denotes the input source function, it is often desired to infer PDE parameters $\theta$, and to estimate its approximate solution from the noisy data. In this scenario, the physics loss is given as $\mathcal{L}_\text{phy}(\beta,\theta)=\frac{1}{2N_c}||\mathcal{G}(\hat{u};\beta,\theta)(x^c,t^c)-f^c||^2_2$ where $f^c=[f(x^c_1,t^c_1),f(x^c_2,t^c_2),...,f(x^c_{N_c},t^c_{N_c})]^{\text{T}}$. Besides, the weight of $\ell_1$ loss $\mu$ in Eq.\eqref{eq:unconstrained optimization} equals zero. 
\subsubsection{Kuramoto-Sivashinsky equation}\label{sec:KS para_est}
\begin{figure}[H]
\centering 
\includegraphics[width=0.85\textwidth]{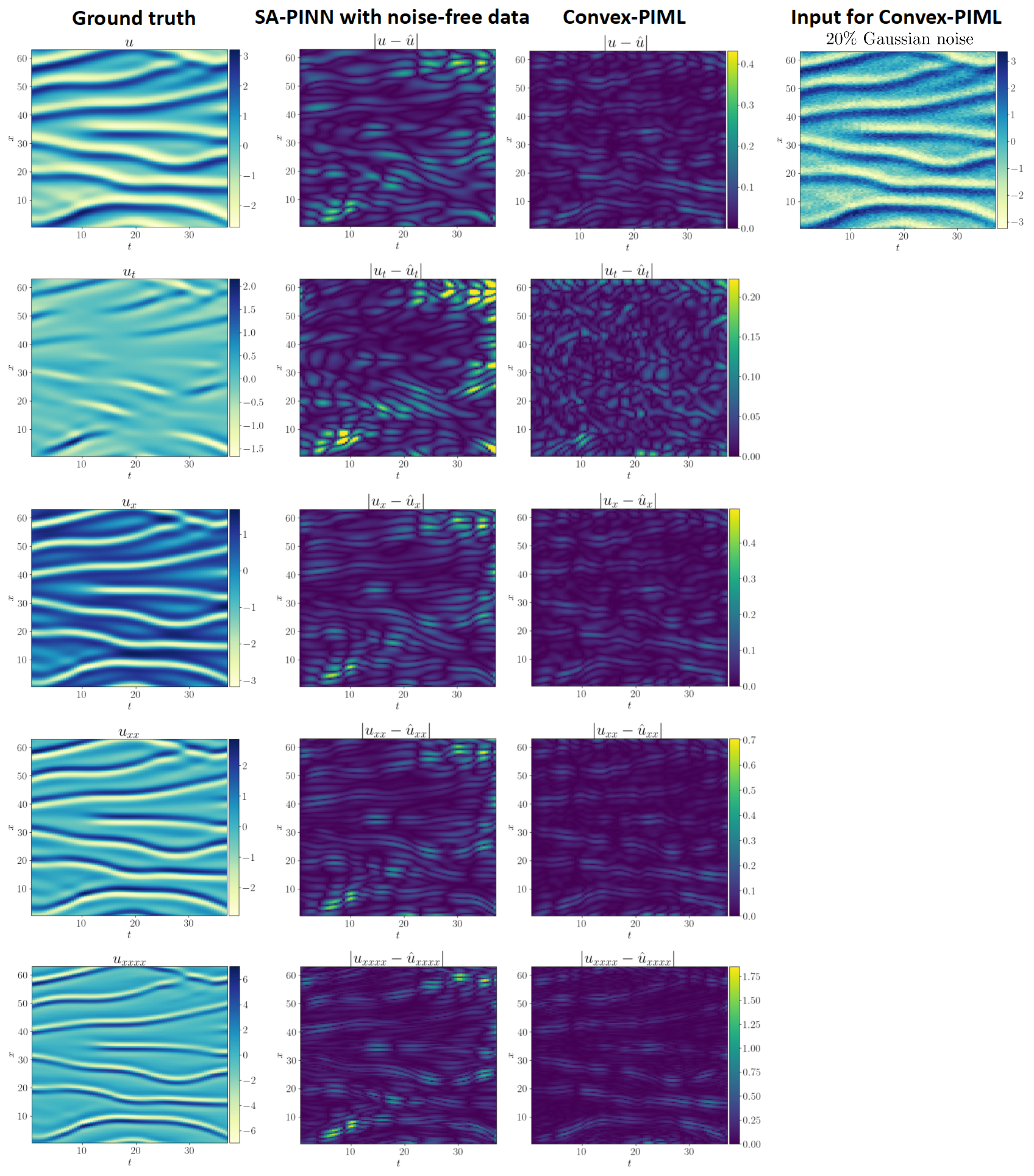}
\caption{The comparisons of estimated solution and hidden physics results of K-S equation between Convex-PIML and SA-PINN. Note that SA-PINN utilizes noise-free data, whereas Convex-PIML operates with noisy data. We choose the best result of SA-PINN among 5 training runs to compare, where estimated parameters $\hat{\theta}=(0.89,0.88,0.89)$.}
\label{fig:real data_KS}
\end{figure}

The Kuramoto–Sivashinsky (K-S) equation, a nonlinear PDE known for its multi-scale dynamics, models phenomena such as concentration waves, flame fronts, and surface flows \cite{smyrlis1991predicting, kevrekidis1990back, lippe2024pde}. It is given by:
\begin{equation}\label{eq:KS}
    u_t + \theta_1u u_x + \theta_2u_{xx} + \theta_3 u_{xxxx} = 0, x \in [0, L_x], t \in [0, L_t],
\end{equation}
with $\theta_1 = \theta_2 = \theta_3 = 1$, $L_x = 63.5$, and $L_t = 37$. The dataset $\mathbf{y} \in \mathbb{R}^{128\times75}$, generated via \cite{brandstetter2022lie} and shown in Fig.~\ref{fig:real data_KS}, displays strong multi-scale characteristics.

To approximate the solution, we initialize B-splines of degree 20 in space and 16 in time, distributing $50 \times 25$ knots uniformly. As uniform knots may be suboptimal, we perform 10 rounds of knot optimization using the data fitting loss $\mathcal{L}_\text{data}$, resulting in improved initialization (see Fig.~\ref{fig:framework}).

\begin{figure}[H]
\centering 
\subfigure[The optimization results of parameter estimation of K-S equation using Convex-PIML with noise-free data.]{
\label{fig:op_results_KS_para_est_noise_free}
\includegraphics[width=0.78\textwidth]{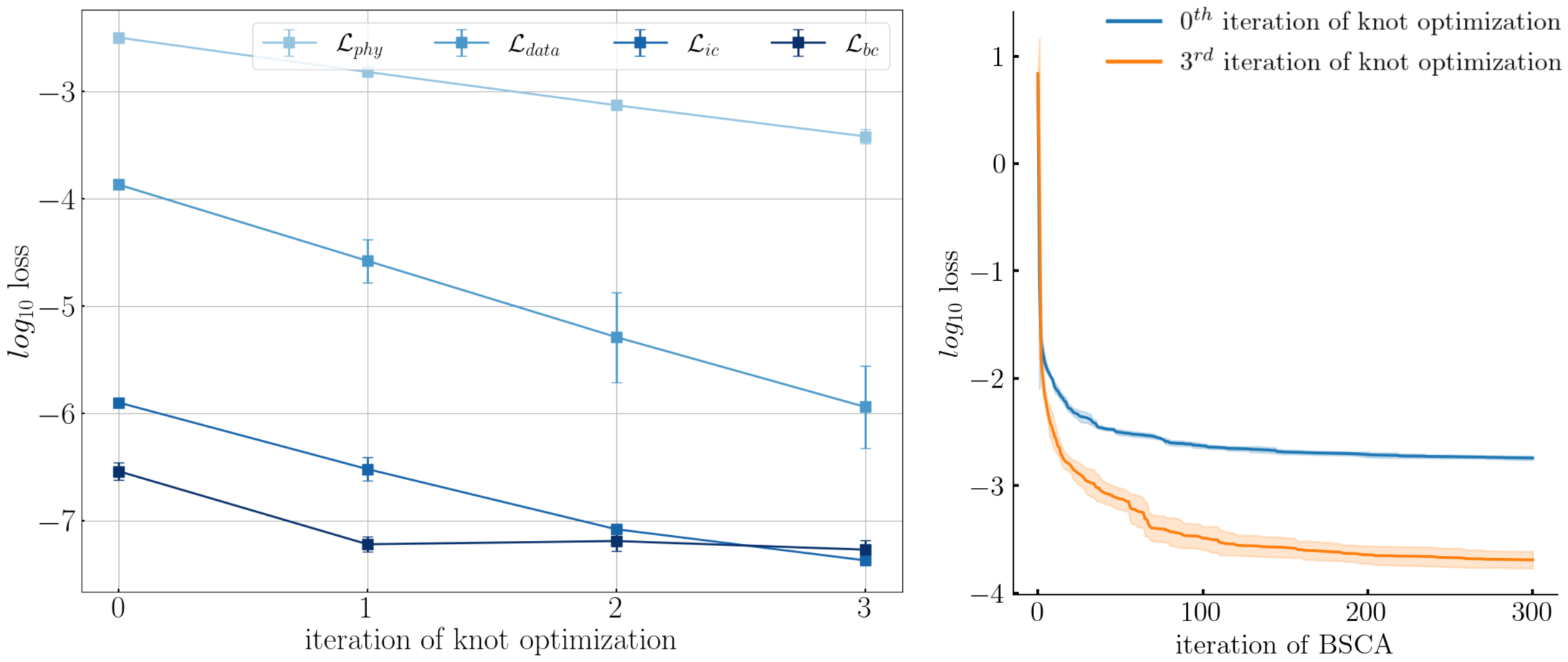}}
\subfigure[The comparisons of optimization results between Convex-PIML and SA-PINN in the parameter estimation of K-S equation using data with different levels of noise. The data fitting loss $\mathcal{L}_\text{data}$ is evaluated with noise-free data.]{
\label{fig:op_results_KS_para_est_compara}
\includegraphics[width=0.78\textwidth]{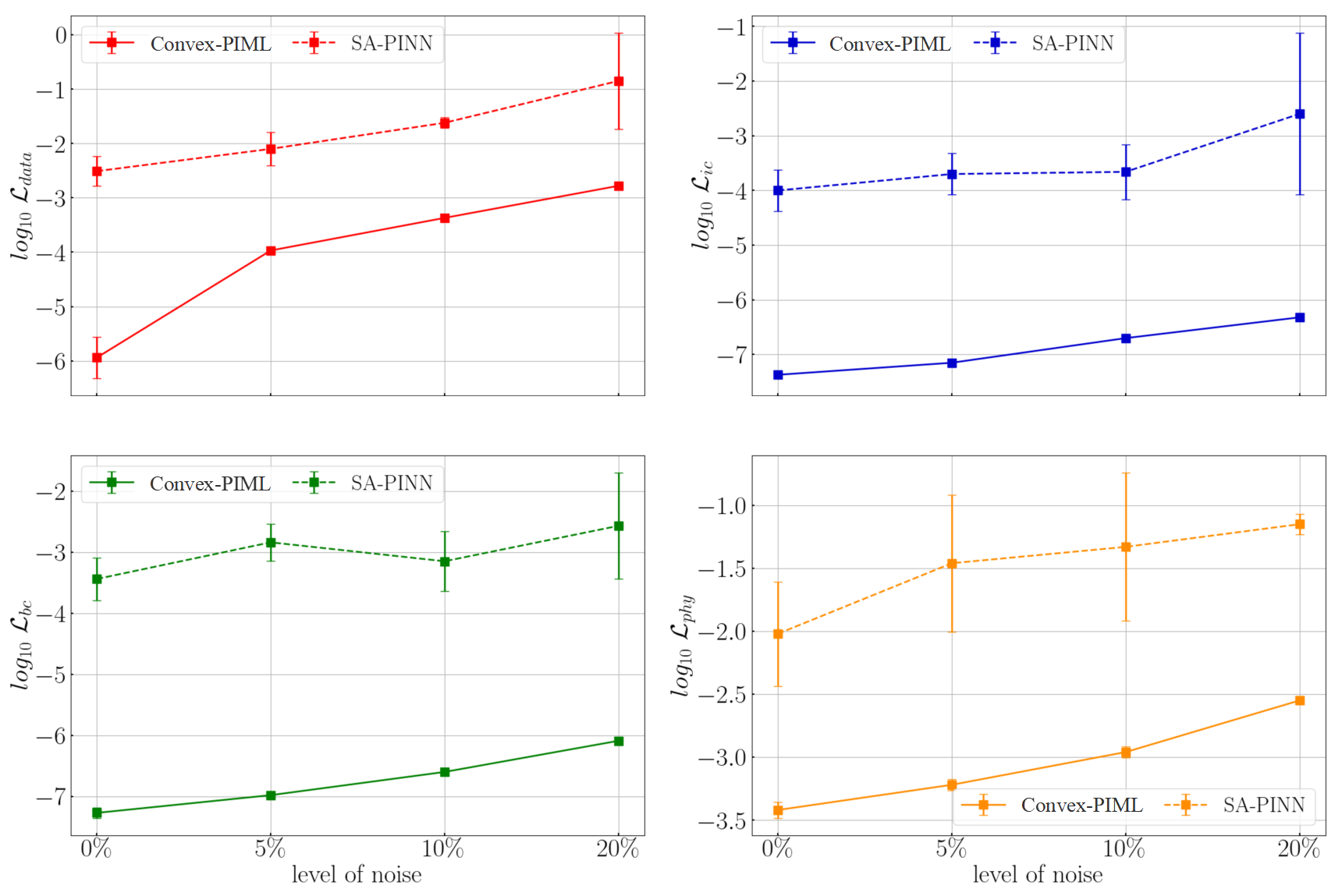}}
\subfigure[The comparisons of parameter estimation results between Convex-PIML and SA-PINN of K-S equation using data with different levels noise.]{
\label{fig:para_results_KS_para_est_compara}
\includegraphics[width=0.78\textwidth]{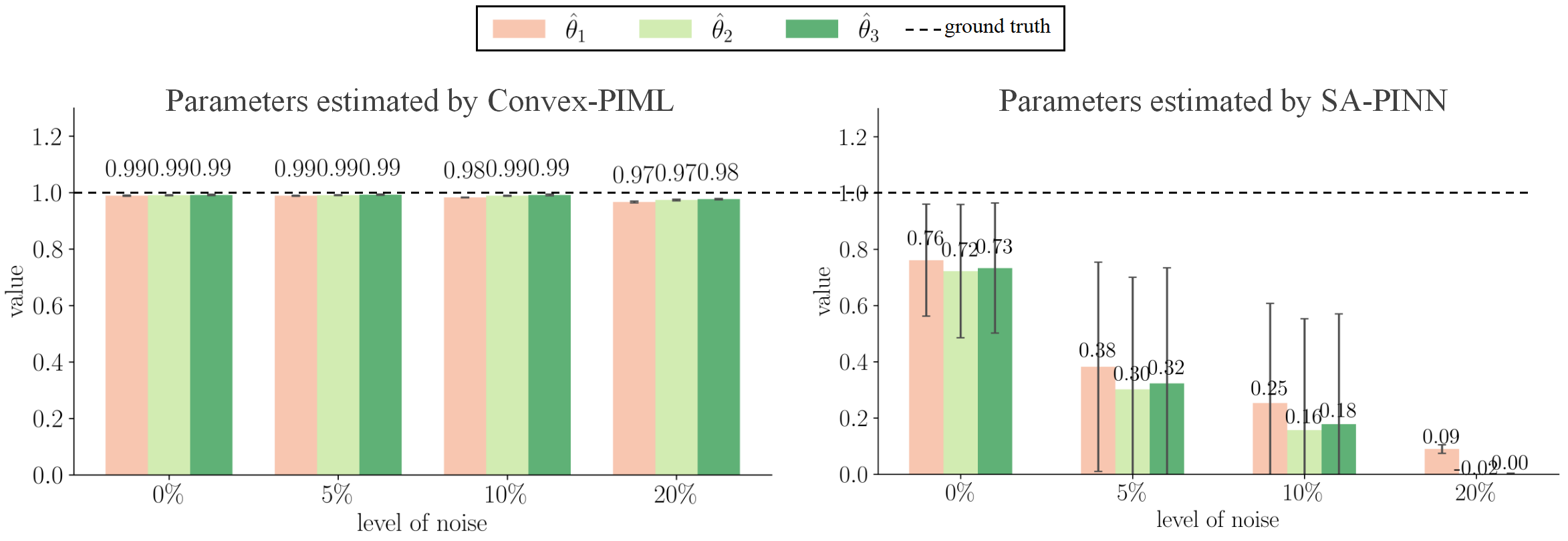}}
\caption{The results of Convex-PIML and the comparisons to SA-PINN \cite{mcclenny2023self} in the parameter estimation of K-S equation. Mean ± 1 standard deviation; 5 training runs.}
\label{fig:results_KS_para_est}
\end{figure}

The optimization problem defined in Eq.\eqref{eq:unconstrained optimization} is solved by executing 300 iterations of BSCA, with step sizes determined through exact line search to estimate the parameters of the K-S equation. The derivation of exact line search for parameter estimation of K-S equation is provided in Appendix \ref{appendix:derivation of exact line search}. During this process, three iterations of knot optimization are performed to enhance the accuracy of both the optimization and parameter estimation (both the data fitting error and physics error distributions are used here). To demonstrate the robustness of Convex-PIML with respect to initialization, we repeat this procedure five times with random initializations of $\beta$ and $\theta$ while keeping the trade-off parameters fixed. The optimization results are presented in Fig.\ref{fig:op_results_KS_para_est_noise_free}.

Notably, all loss terms significantly decrease with the knot optimization, confirming the effectiveness of this approach in mitigating spectral bias. Furthermore, the loss curves exhibit small standard deviations, indicating that Convex-PIML achieves nearly identical accuracy across different initializations. The loss curves on the right side also suggest this point. This arises from BSCA leveraging the weakly non-convex structure of our defined PIML optimization problem, thereby inheriting the advantage of convex optimization: the existence of a unique optimal. 

To further demonstrate the robustness of Convex-PIML to noise, we introduce Gaussian noise levels of $5\%$, $10\%$, and $20\%$ into the data. We then repeat the optimization procedure with these noisy datasets, and compare Convex-PIML to the self-adaptive physics-informed neural network (SA-PINN) \cite{mcclenny2023self}. In SA-PINN, a function linking error to weight is predefined for both data and collocation points, allowing the adaptive optimization of point weights based on gradients derived from this function. Specifically, weights for points with larger errors are increased, enabling SA-PINN to concentrate on harder-to-train regions. We perform 10,000 iterations of Adam optimization to minimize data fitting loss, followed by 50,000 iterations of L-BFGS to minimize both data fitting loss and physics loss, following the optimization strategy in \cite{mcclenny2023self}. For this comparative experiment, SA-PINN is also repeatedly trained five times with different initializations.

The optimization results for Convex-PIML and SA-PINN, using noisy datasets, are presented in Fig.\ref{fig:op_results_KS_para_est_compara}. Although the accuracy of Convex-PIML does decrease with increasing noise intensity, it consistently outperforms SA-PINN. Furthermore, due to the sensitivity of non-convex optimization to initialization, the results from SA-PINN exhibit considerable variability across different initializations, resulting in larger error bars for losses, particularly for physics loss. Besides, SA-PINN exhibits a pronounced sensitivity to noise, with a marked decline in the precision of parameter estimation observed as noise levels escalate, as demonstrated in Fig.\ref{fig:para_results_KS_para_est_compara}. By comparison, Convex-PIML yields parameter estimation results that are significantly more stable and closer to the ground truth.

We also compare the hidden physics inferred by SA-PINN using noise-free data and Convex-PIML using data with $20\%$ Gaussian noise. We select the best result of SA-PINN for comparison, yielding estimated parameters of $\hat{\theta} = (0.89,0.88,0.89)$. As shown in Fig.\ref{fig:real data_KS}, even using noise-free data, SA-PINN still produces many regions with high errors, as the user-defined function does not always accurately reflect the relationship between error and weight. In contrast, the hidden physics quantities inferred by Convex-PIML using noisy data are much closer to the ground truth, thanks to our proposed knot optimization method, which theoretically guarantees error reduction. Notably, the accuracy of Convex-PIML can be further enhanced through continuous knot optimization.

\subsubsection{Navier-Stokes equation}\label{sec:NS para_est}
The Navier–Stokes (N-S) equations are a set of coupled, nonlinear partial differential equations (PDEs) that govern the dynamics of fluid motion. These equations are fundamental across a broad spectrum of scientific disciplines, with applications ranging from climate modeling and analysis of blood flow in the human body to the study of ocean currents and pollution dispersion, among others. In this particular study, we examine the case of incompressible flow around a cylinder, which results in an asymmetric vortex shedding pattern in the cylinder’s wake. The mathematical formulation of the problem, expressed in terms of vorticity field: $\omega$, and velocity fields: $u$ and $v$, is represented by:
\begin{equation}\label{eq:NS}
     \omega_t  + \theta_1 u\omega_{x} + \theta_2 v\omega_{y} + \theta_3\omega_{xx} + \theta_4\omega_{yy}=0, x \in [0, L_x], y \in [0, L_y], t \in [0, L_t]
\end{equation}
The two components of the velocity field data $u(x,y,t)$ and $v(x,y,t)$ are obtained from \cite{raissi2019physics} where the numerical solution of Eq.\eqref{eq:NS} is performed for the parameters $\theta=(1.0,1.0,-0.01.-0.01)$. The two components of velocity field datasets $(u,v)\in\mathbb{R}^{100\times50\times20}$ are perturbed $5\%$, $10\%$, and $20\%$ Gaussian noise to simulate the measured data. The discrete measurement data is acquired over a rectangular domain of $x\in[1.0,8.0]$ and $y\in[-2.0,2.0]$ with the period of $t\in[0,1.9]$.

We begin by initializing the B-splines, setting their degrees to 7 in the $x$-dimension, 7 in the $y$-dimension, and 4 in the $t$-dimension. This setup ensures a smooth approximation of the solution. We uniformly distribute $10 \times 6 \times 4$ knots across the domain defined by $x \in [1.0, 8.0]$, $y \in [-2.0, 2.0]$, and the temporal period $t \in [0, 1.9]$. Similarly, we perform three iterations of knot optimization based on data fitting error distribution to optimize the initial B-splines.

\begin{figure}[H]
\centering 
\subfigure[The loss curves of parameter estimation of N-S equation using Convex-PIML with noisy data. The data fitting loss is evaluated on data with corresponding noise. The spike features observed in the loss curves are attributed to the proposed mini-batch BSCA.]{
\label{fig:op_results_NS_para_est_noise_free}
\includegraphics[width=0.72\textwidth]{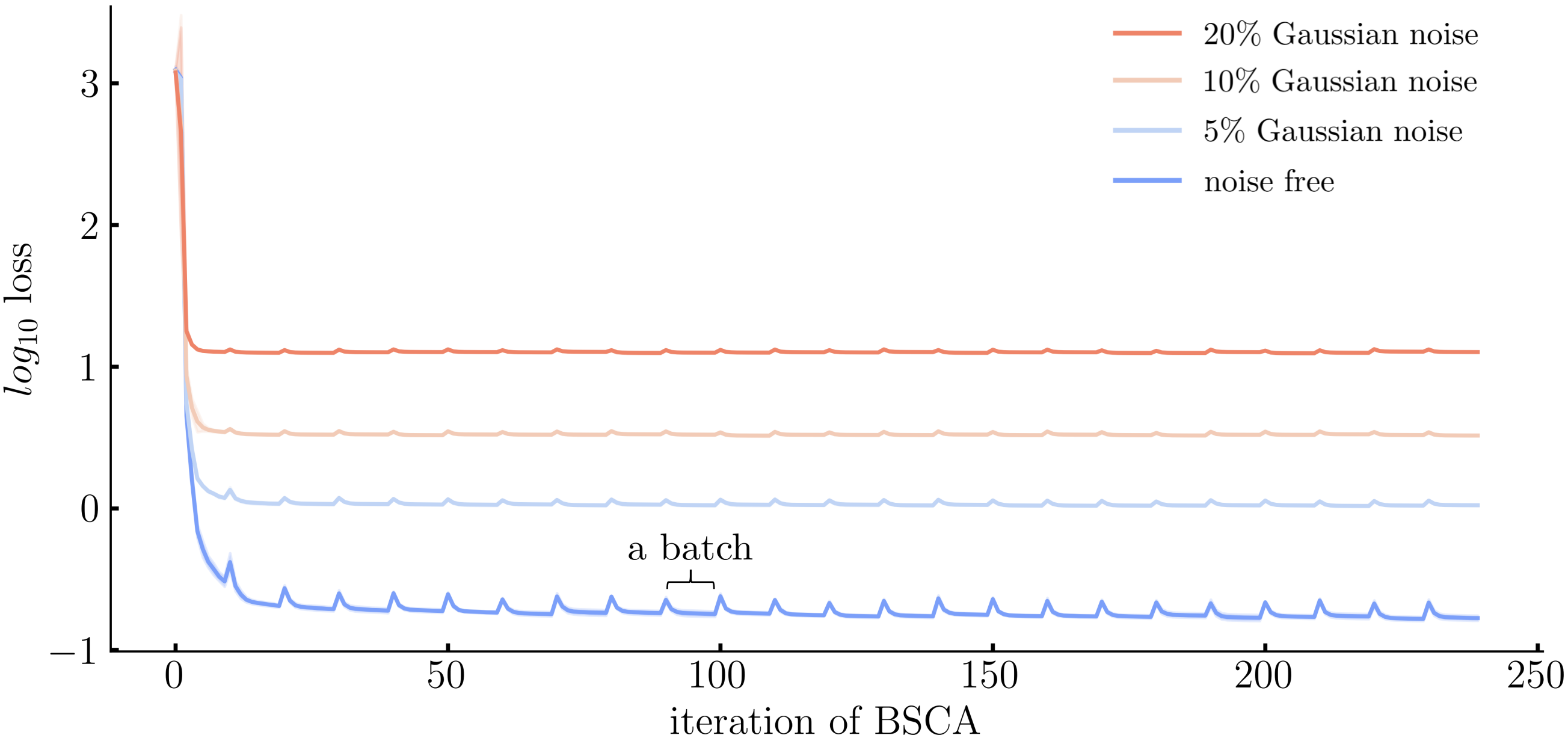}}
\subfigure[The comparisons of parameter estimation results between Convex-PIML and SA-PINN of N-S equation using data with different levels noise.]{
\label{fig:para_results_NS_para_est_compara}
\includegraphics[width=0.72\textwidth]{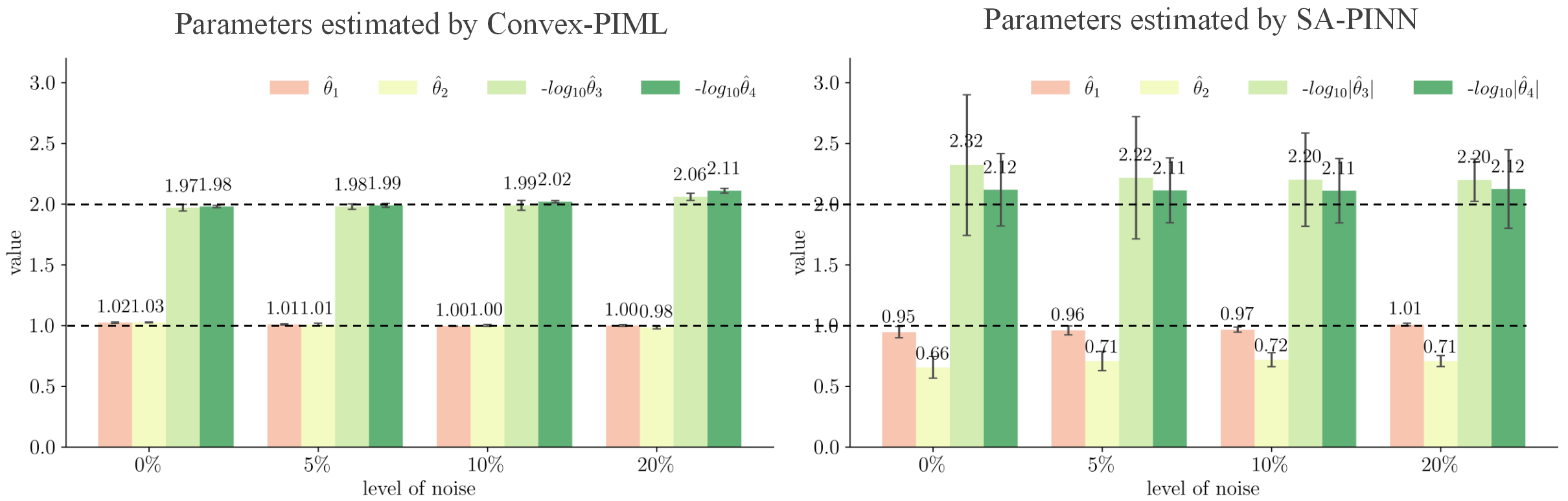}}
\subfigure[The comparisons of approximated solution ($t=1 s$) of N-S equation between
Convex-PIML and SA-PINN. We choose the best result of SA-PINN among 5 training runs to compare,
where estimated parameters $\theta = (1.0274,0.7593,-0.0304,0.0281)$.]{
\label{fig:hid_results_NS_para_est_20noise}
\includegraphics[width=0.72\textwidth]{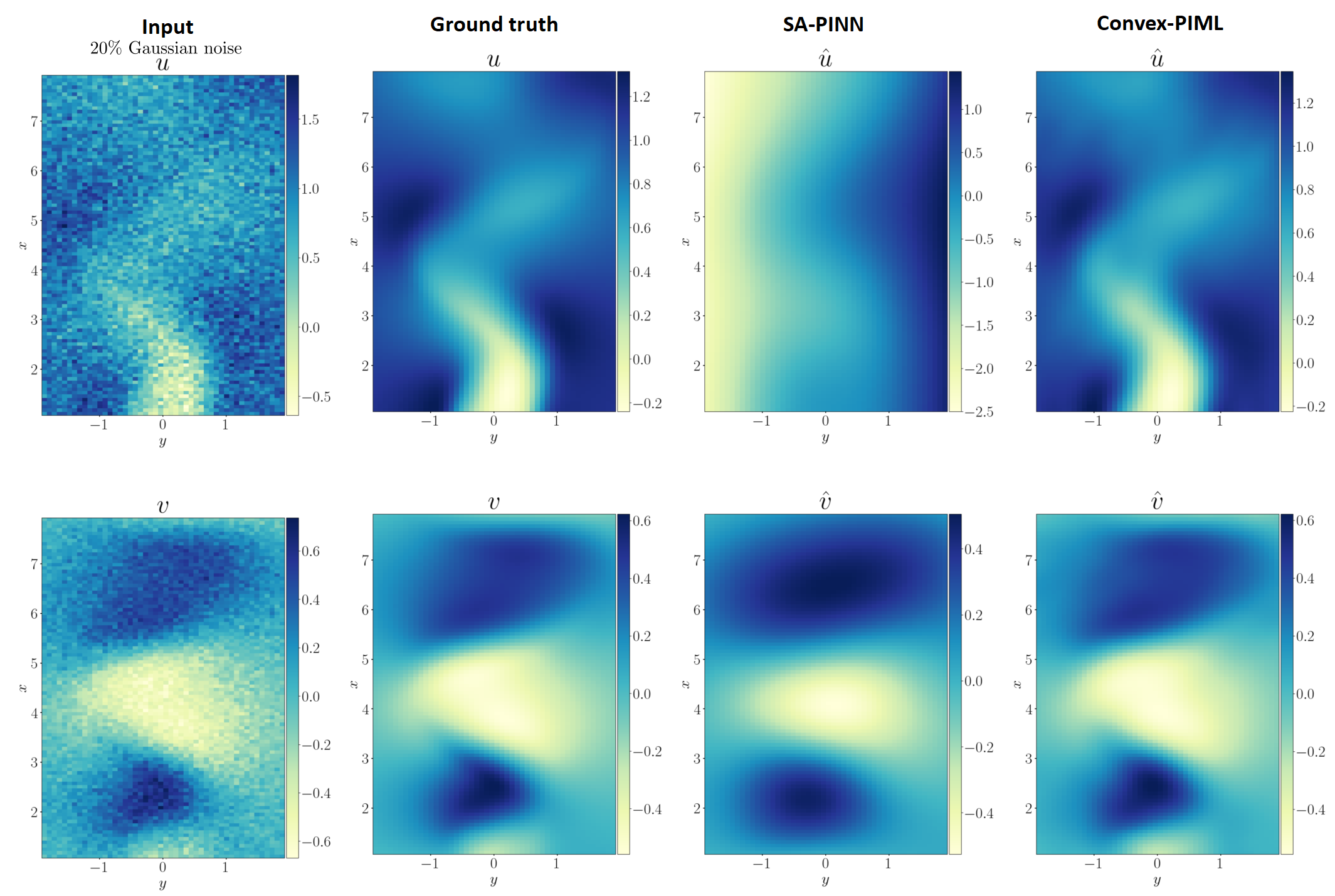}}
\caption{The results of Convex-PIML and the comparisons to SA-PINN \cite{mcclenny2023self} in the parameter estimation of N-S equation. Mean ± 1 standard deviation; 5 training runs.}
\label{fig:results_NS_para_est}
\end{figure}

Inputting all the data into the model results in large matrices, significantly reducing computational efficiency and potentially leading to memory overflow. To address this, we divide the data into batches and input them sequentially to generate much smaller matrices, which we call mini-batch BSCA. The optimization problem, defined in Eq.\eqref{eq:unconstrained optimization}, is then solved using mini-batch BSCA. Specifically, we perform four epochs; in each epoch, the dataset is randomly divided into six batches, and ten iterations of BSCA are conducted on each batch. The step sizes are determined through exact line search, with the derivation process provided in Appendix \ref{appendix:derivation of exact line search}. Since the Navier–Stokes equation data does not exhibit multi-scale features, we find that the optimization results are satisfactory without the need for further knot optimization. Similarly, we repeat the procedure five times with random initializations, while keeping the trade-off parameters constant. We also compare the results of Convex-PIML to SA-PINN. The results are presented in Fig.\ref{fig:results_NS_para_est}.

The loss curves depicted in Fig.\ref{fig:op_results_NS_para_est_noise_free} demonstrate rapid convergence, underscoring the effectiveness of the mini-batch BSCA. This suggests that Convex-PIML is well-suited for handling large datasets. The approximated solutions at $t=1s$, obtained using Convex-PIML and SA-PINN, are illustrated in Fig.\ref{fig:hid_results_NS_para_est_20noise}. Despite the N-S equation data lacking multi-scale features, SA-PINN still fails to accurately approximate $u$. This failure is attributed to SA-PINN's inability to address multi-objective optimization challenges, leading to a significantly poorer fit for $u$ compared to $v$. Consequently, the parameter estimation results from SA-PINN are inaccurate, with the viscosity coefficient estimation potentially yielding negative values. In contrast, Convex-PIML provides an accurate approximation of the true solution, resulting in stable and precise parameter estimation.

\subsection{Equation discovery with partially known equations}

In some cases, only partial (or even no) terms of a governing equation are known, and the goal is to discover the missing terms from the data. Sparse Identification of Nonlinear Dynamics (SINDy) \cite{brunton2016discovering} is a pioneering framework for such equation discovery. This method employs finite-difference approximations of derivatives from time-series data to construct a library of candidate functions. It then uses sequential threshold least squares to identify a parsimonious representation of the system's dynamics. Here, we draw inspiration from SINDy's approach to building the library; specifically, the candidate governing equation is represented as $u_t = \mathbf{\Phi} \theta$, where $\mathbf{\Phi} = \mathbf{\Phi}(u) \in \mathbb{R}^{s}$ is an extensive library of symbolic functions composed of numerous candidate terms organized into a vector. For example, $\mathbf{\Phi} = [1, u, u \circ u, \ldots, u_x, \ldots, \sin(u), \ldots, u \circ u_x, \ldots]^T$. In this context, the physics loss is expressed as $\mathcal{L}_\text{phy}(\beta, \theta) = \frac{1}{2N_c} ||\hat{u}_t(\beta) - \mathbf{\Phi}\big(\hat{u}(\beta)\big) \theta||_2^2$. Unlike the sequential threshold least squares approach used in \cite{brunton2016discovering} to get sparse solutions, we directly solve the Least Absolute Shrinkage and Selection Operator (LASSO) problem to determine the weights of the candidate functions in the library, i.e., the weight of the $\ell_1$ loss term $\mu$ in Eq.\eqref{eq:unconstrained optimization} is non-zero, thereby promoting sparsity more effectively.

\subsubsection{Kuramoto-Sivashinsky equation}\label{sec:KS SINDy}
In this section, we utilize the same dataset as described in Section \ref{sec:KS para_est} to discover the 1D K-S equation as shown in Eq.\eqref{eq:KS}. We employ a total of 35 candidate functions to construct the SINDy library. The setting and optimization strategy for the initial B-splines is consistent with that in Section \ref{sec:KS para_est}. Subsequently, we solve the optimization problem defined in Eq.\eqref{eq:unconstrained optimization} to determine the weight of each candidate function within the constructed library. The diminishing step size presented in Eq.\eqref{eq:diminishing step size} is used to update $\beta$, as the computational burden of performing an exact line search is high due to the large number of candidate functions. Drawing inspiration from the sequential threshold least squares method introduced in \cite{brunton2016discovering}, we propose the Sequential Threshold Least Absolute Shrinkage and Selection Operator (ST-LASSO). The LASSO problem of updating $\theta$ in Eq.\eqref{eq:update theta} is solved by Fast Iterative Shrinkage-Thresholding Algorithm (FISTA) \cite{beck2009fast}. Additionally, we apply a threshold of 0.2 to eliminate candidate equations with weights below this threshold from the library after solving the PIML optimization problem. This process is repeated with the updated library until no candidate functions have weights below the threshold. Ultimately, we derive the specific form of the governing equation using the remaining candidate functions. 

\begin{figure}[H]
\centering 
\subfigure[The equation discovery result of K-S equation using noise-free data.]{
\label{fig:op_results_KS_eq_dis_noise_free}
\includegraphics[width=0.85\textwidth]{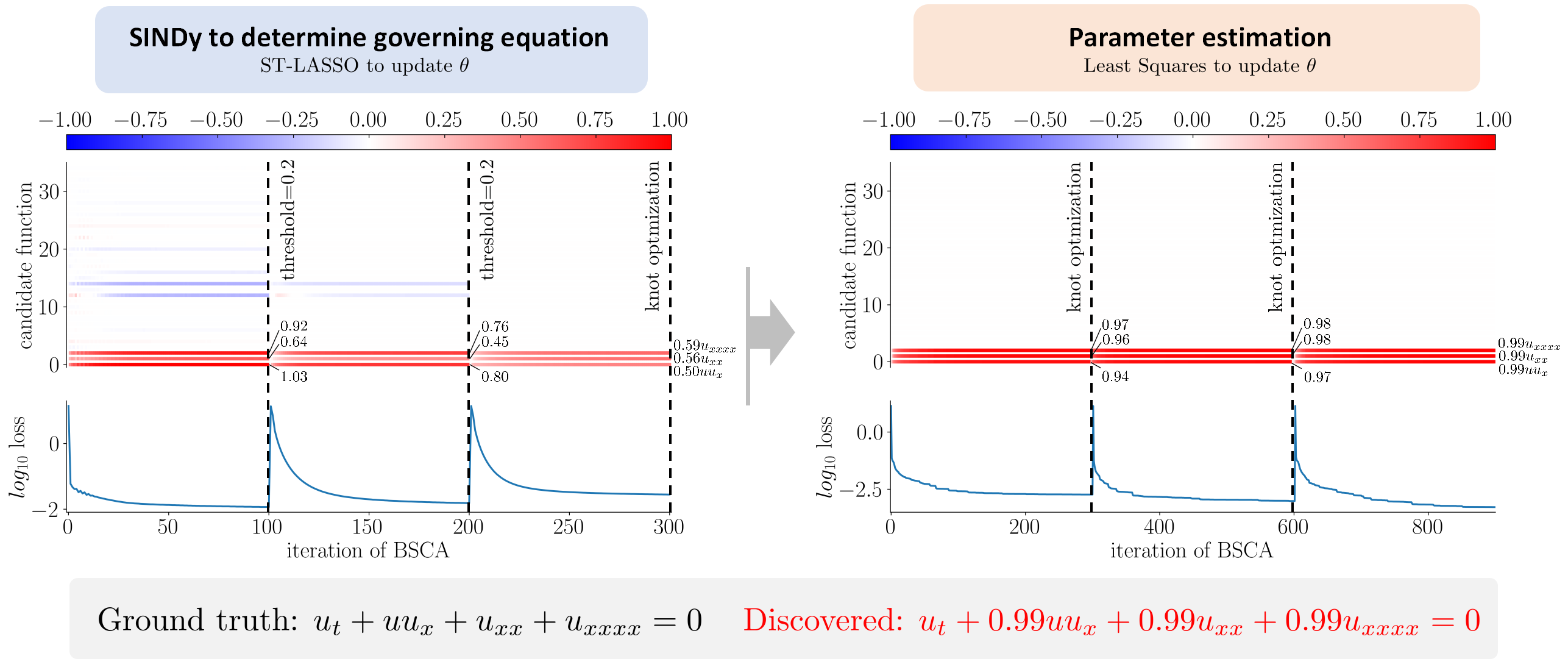}}
\subfigure[The equation discovery result of K-S equation using data with $20\%$ Gaussian noise.]{
\label{fig:op_results_KS_eq_dis_20noise}
\includegraphics[width=0.85\textwidth]{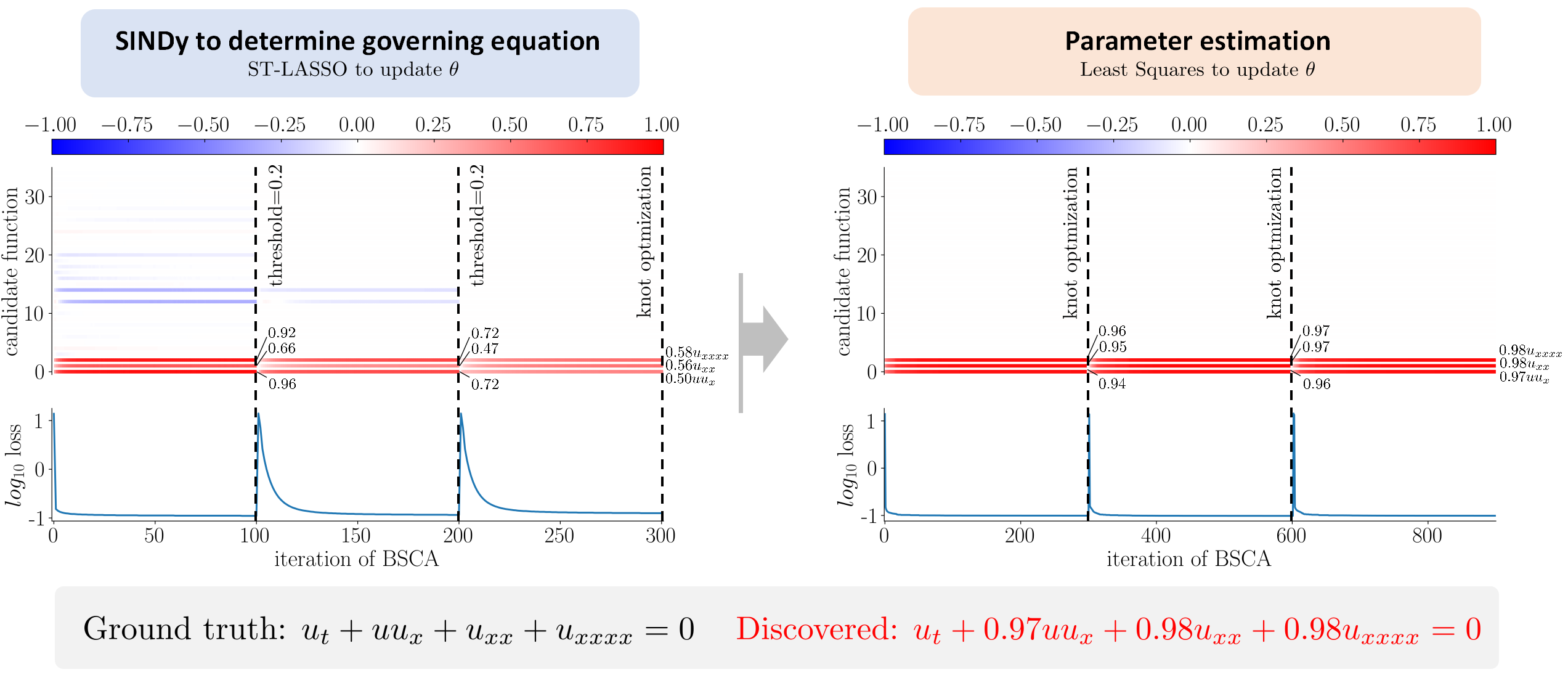}}
\subfigure[The estimated hidden physics using data with $20\%$ Gaussian noise after the first 100 iterations of BSCA in the SINDy to determine governing equation.]{
\label{fig:hid_results_KS_eq_dis_20noise}
\includegraphics[width=0.85\textwidth]{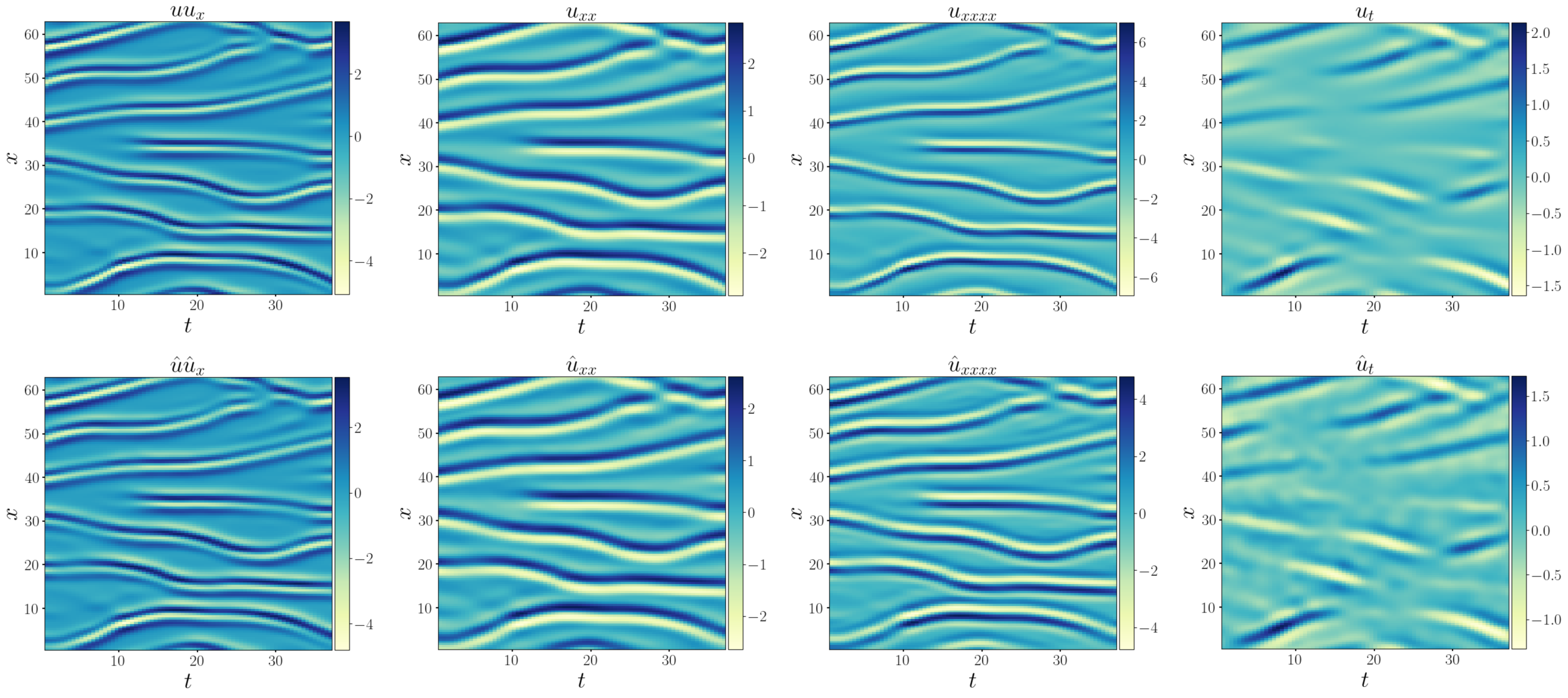}}
\caption{The equation discovery result of K-S equation using data with different level of noise, where ST-LASSO refers to Sequential Threshold Least Absolute Shrinkage and Selection Operator.}
\label{fig:results_KS_eq_dis}
\end{figure}

The above procedure of determining the candidate functions of governing equation is referred as \textit{SINDy to determine governing equation}. Due to the regularization of the $\ell_1$ loss, the physics loss cannot be further minimized, which leads to reduced accuracy in parameter estimation. Consequently, we set $\mu=0$, transforming the task into a parameter estimation problem. We then follow the procedure outlined in Section \ref{sec:KS para_est} to estimate the accurate parameters of the discovered equation, especially the knot optimization to mitigate the spectral bias. This methodology is applied to both noise-free data and data with $20\%$ Gaussian noise. The results of the equation discovery process are depicted in Fig.\ref{fig:results_KS_eq_dis}.

The optimization process, as depicted in the loss curves in Figs.\ref{fig:op_results_KS_eq_dis_noise_free} and \ref{fig:op_results_KS_eq_dis_20noise}, converges rapidly. The hidden physics estimated from data with $20\%$ Gaussian noise after 100 iterations of BSCA in the SINDy to determine the governing equation, are presented in Fig.\ref{fig:hid_results_KS_eq_dis_20noise}. These results show a strong agreement with the ground truth, except for $\hat{u}_{xxxx}$. Although the peak values of $\hat{u}_{xxxx}$ are lower than $u_{xxxx}$ due to spectral bias, the approximation still accurately capture the essential features of ground truth. It indicates that Convex-PIML can quickly obtain an well-approximated solution $\hat{u}(x, t)$ due to the inherent advantages of convex optimization in Convex-PIML. Consequently, the weights of correct candidate functions increase rapidly during optimization. The LASSO problem for updating $\theta$ and enhancing sparsity is efficiently solved using FISTA, swiftly reducing the weights of most of the incorrect candidate functions to zero. However, the inaccurate estimation of $u_{xxxx}$ results in some incorrect candidate functions retaining non-zero weights. To address this, we employ the proposed ST-LASSO, which further enhances sparsity by eliminating candidate functions with weights below a specified threshold. To avoid excluding correct candidate functions, this threshold is generally set to a low value; in this case, it is set at 0.2. Conversely, the weight of the $\ell_1$ loss $\mu$ is set high, ensuring that the weights of most incorrect candidate functions are reduced to zero or near zero. Compared to sequential threshold least squares, ST-LASSO accelerates the elimination of incorrect candidate functions. As shown in Figs. \ref{fig:op_results_KS_eq_dis_noise_free} and \ref{fig:op_results_KS_eq_dis_20noise}, this strategy guarantees the accurate discovery of correct candidate functions. Subsequently, parameter estimation is performed to obtain accurate parameters of the discovered governing equation. A comparison of results from both noise-free and noisy datasets reveals that the discovered equations are nearly identical, underscoring the robustness of Convex-PIML against noise interference.

\subsubsection{Navier-Stokes equation}\label{sec:NS SINDy}

\begin{figure}[H]
\centering 
\subfigure[The equation discovery result of N-S equation using noise-free data.]{
\label{fig:op_results_NS_eq_dis_noise_free}
\includegraphics[width=0.82\textwidth]{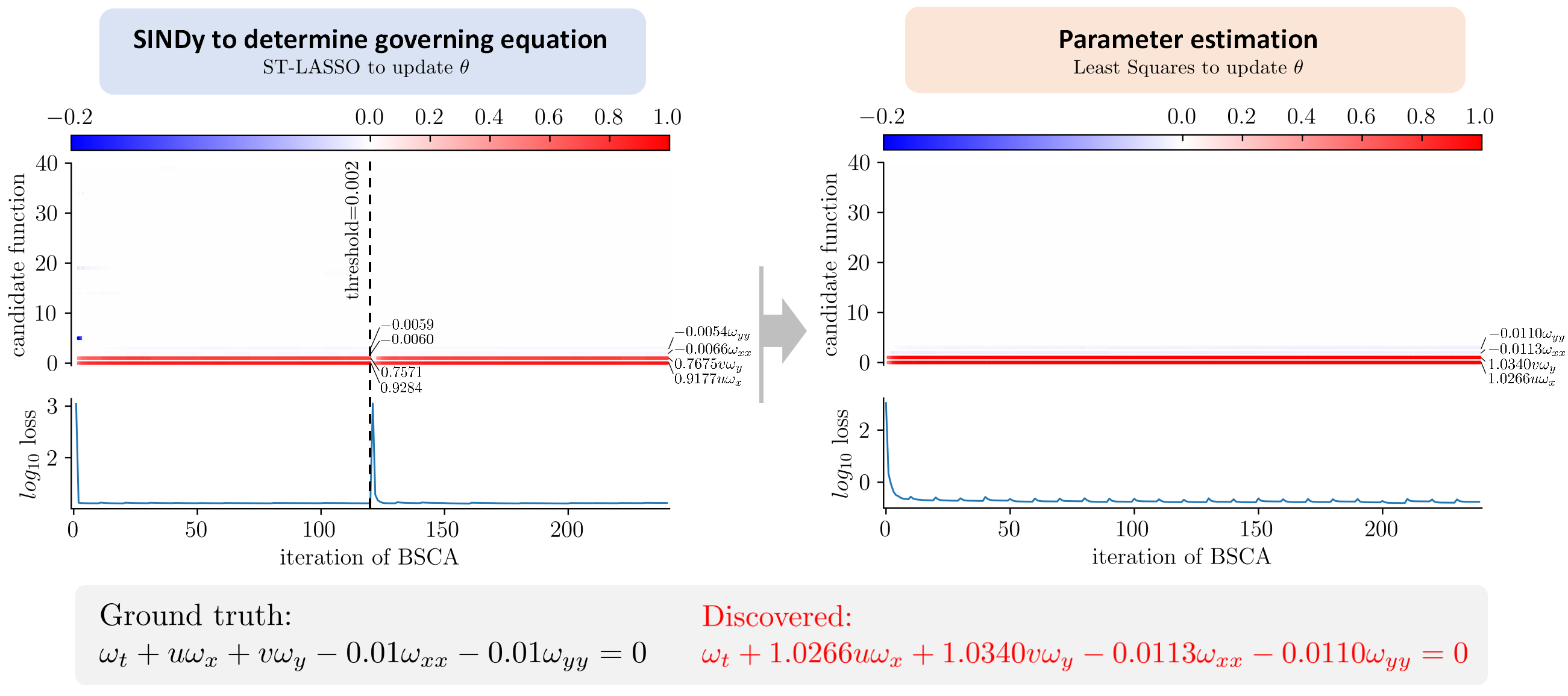}}
\subfigure[The equation discovery result of N-S equation using data with $20\%$ Gaussian noise.]{
\label{fig:op_results_NS_eq_dis_20noise}
\includegraphics[width=0.82\textwidth]{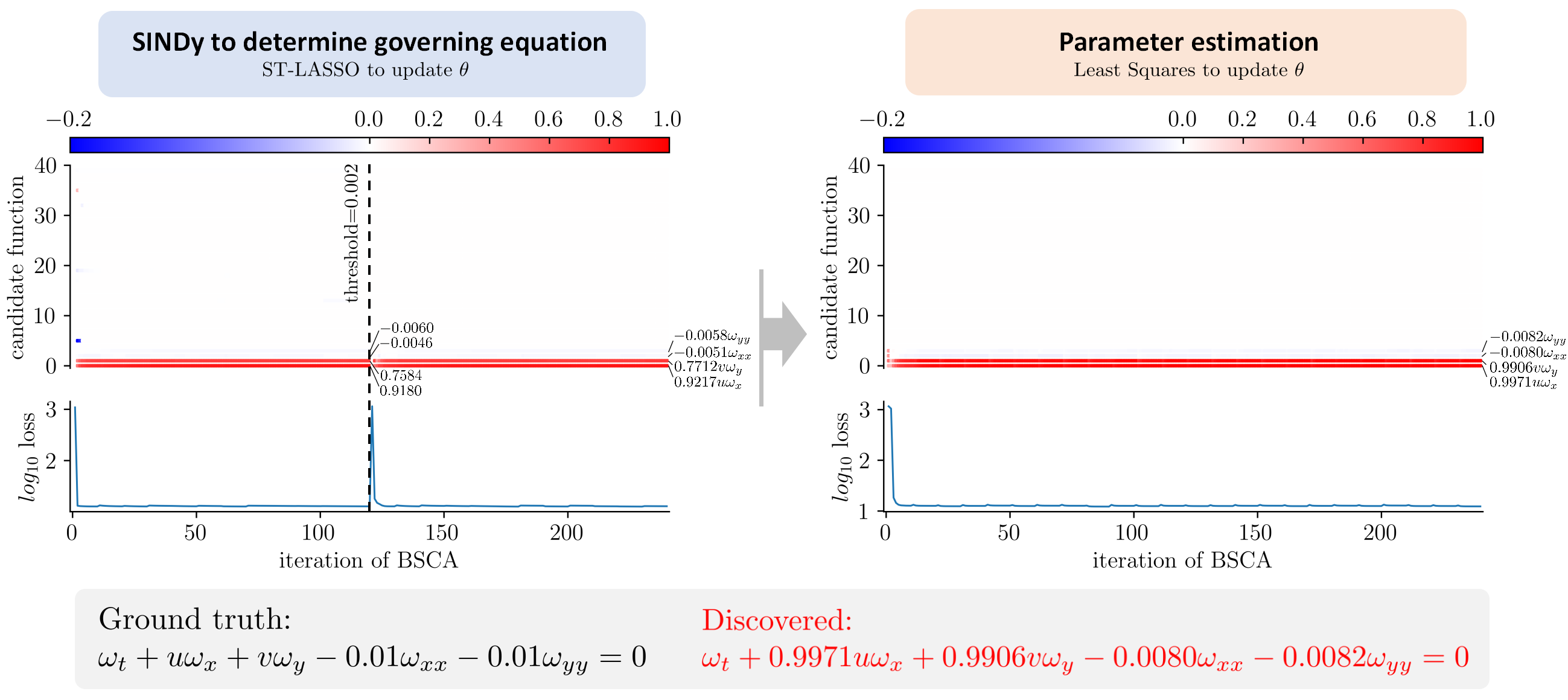}}
\subfigure[The estimated hidden physics ($t=1 s$) using data with $20\%$ Gaussian noise after the first 120 iterations of BSCA in the SINDy to determine governing equation.]{
\label{fig:hid_results_NS_eq_dis_20noise}
\includegraphics[width=0.82\textwidth]{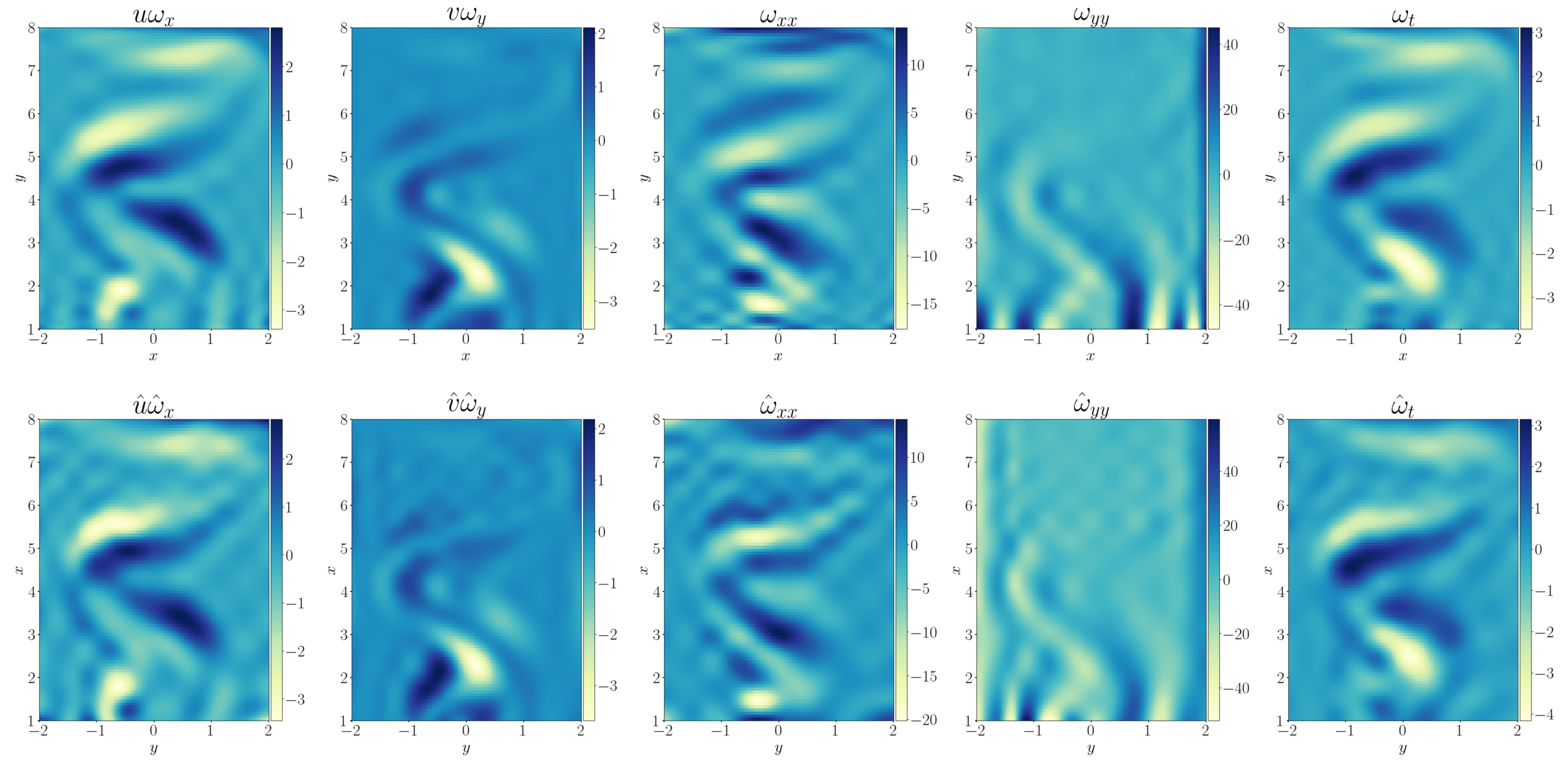}}
\caption{The equation discovery result of N-S equation using data with different levels of noise, where ST-LASSO refers to Sequential Threshold Least Absolute Shrinkage and Selection Operator.}
\label{fig:results_NS_eq_dis}
\end{figure}

In this section, we utilize the same dataset described in Section \ref{sec:NS para_est} to identify the 2D N-S equation in Eq.\eqref{eq:NS}. A total of 40 candidate functions are employed to construct the SINDy library. The initial B-splines settings and optimization strategy align with those outlined in Section \ref{sec:NS para_est}. We then address the optimization problem formulated in Eq.\eqref{eq:unconstrained optimization} using the diminishing step size and mini-batch BSCA to determine the weights of each candidate function within the library. Specifically, the process involves two epochs, each consisting of the dataset being randomly divided into six batches, followed by ten iterations of BSCA per batch. Subsequent parameter estimation is performed with the $\ell_1$ loss weight $\mu$ set to zero. The results from the equation discovery process using both noise-free data and data with $20\%$ Gaussian noise are illustrated in Fig.\ref{fig:results_NS_eq_dis}.

In contrast to the K-S equation in Eq.\eqref{eq:KS}, discovering the N-S equation poses significant challenges due to the small viscosity coefficient, which is only 0.01. This small coefficient increases the risk of incorrectly excluding $\omega_{xx}$ and $\omega_{yy}$ due to a too large threshold. To circumvent this issue, the following criteria must be satisfied during the equation discovery process: (1) ensuring that the weights of the correct candidate functions closely approximate their true values and (2) minimizing the weights of incorrect candidate functions to approach zero. Our proposed method effectively addresses these challenges. By leveraging the strengths of convex optimization, we can rapidly obtain a good approximate solution $\hat{u}(x,t)$, as illustrated in Fig.\ref{fig:hid_results_NS_eq_dis_20noise}, which facilitates the quick convergence of the weights of the correct candidate functions to their true values. Additionally, by assigning a substantial weight to the $\ell_1$ loss and efficiently solving the LASSO optimization problem for updating $\theta$ using the FISTA, the weights of all incorrect candidate functions are swiftly reduced to zero or near-zero values, as demonstrated in Figs. \ref{fig:op_results_NS_eq_dis_noise_free} and \ref{fig:op_results_NS_eq_dis_20noise}. Consequently, we can set a very small threshold of 0.002 to filter out incorrect candidate functions while preserving the correct ones after a single iteration of the ST-LASSO. Furthermore, the accuracy of discovering the N-S equations remains robust even when using noisy data.

\subsection{State prediction with the conservation law}
\begin{figure}[H]
\flushleft
\label{fig:double_pendulum_system}
\includegraphics[width=0.8\textwidth]{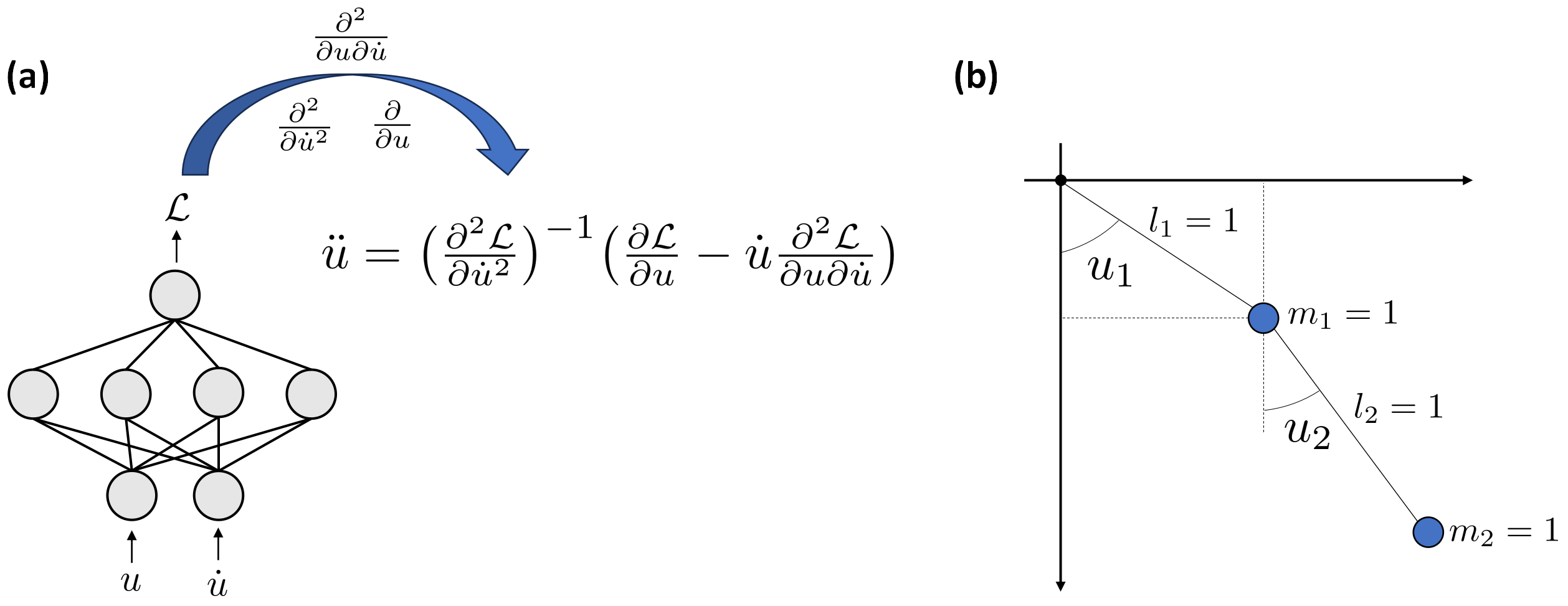}
\caption{(a) The architecture of the Lagrangian neural network and (b) the double pendulum system.}
\end{figure}

In certain scenarios, the governing equations are exceedingly complex, making it impractical to derive them directly from data. The construction of accurate models relies on understanding the underlying symmetries of the system. In physics, these symmetries are often associated with conservation laws, such as those for energy and momentum, which can be elegantly described by the Euler-Lagrange equation: 
\begin{equation}\label{eq:EL}
    \frac{d}{dt}\frac{d\mathcal{L}}{d\dot{u}}-\frac{d\mathcal{L}}{du}=0,
\end{equation}
where $\mathcal{L}$ denotes Lagrangian that enforces conservation of total energy, $z=(u,\dot{u})$ denote the state of a classical physics system. Traditionally, physicists derive an analytical expression for the Lagrangian $\mathcal{L}$ and subsequently expand the Euler-Lagrange equation into a system of differential equations. However, obtaining an analytical form for $\mathcal{L}$ can be challenging. To address this, Cranmer et al. \cite{cranmer2020lagrangian} introduced Lagrangian Neural Networks (LNNs), which learn the Lagrangian directly from data by parameterizing it with a neural network, denoted as $\mathcal{L}(\theta)$. The input to LNN consists of $z=(u, \dot{u})$, and the output is the learned Lagrangian $\mathcal{L}(\theta)$, as shown in Fig.\ref{fig:double_pendulum_system} (a). By re-expressing the Euler-Lagrange equation, the predicted acceleration $\ddot{u}$ can be computed using the learned Lagrangian: $\ddot{\hat{u}} = (\nabla_{\dot{u}}\nabla_{\dot{u}}^\top\mathcal{L}(\theta))^{-1}[\nabla_u\mathcal{L}(\theta) - (\nabla_u\nabla_{\dot{u}}^\top\mathcal{L}(\theta))\dot{u}]$ \cite{cranmer2020lagrangian}. This allows for the integration of $\ddot{\hat{u}}$ to obtain the system's dynamic state, $\hat{z} = (\hat{u}, \dot{\hat{u}})$. The loss function used to train the LNN is defined as the squared discrepancy between the predicted and true states $||\hat{z}(\theta) - z||_2^2$. In this approach, the physics prior is implicitly embedded within the neural network.

\begin{figure}[H]
\centering 
\subfigure[The loss curves of training vanilla LNN and prediction of $\ddot{u}_1$ and $\ddot{u}_2$ on test dataset using noisy data directly.]{
\label{fig:double_pendulum_loss}
\includegraphics[width=1.\textwidth]{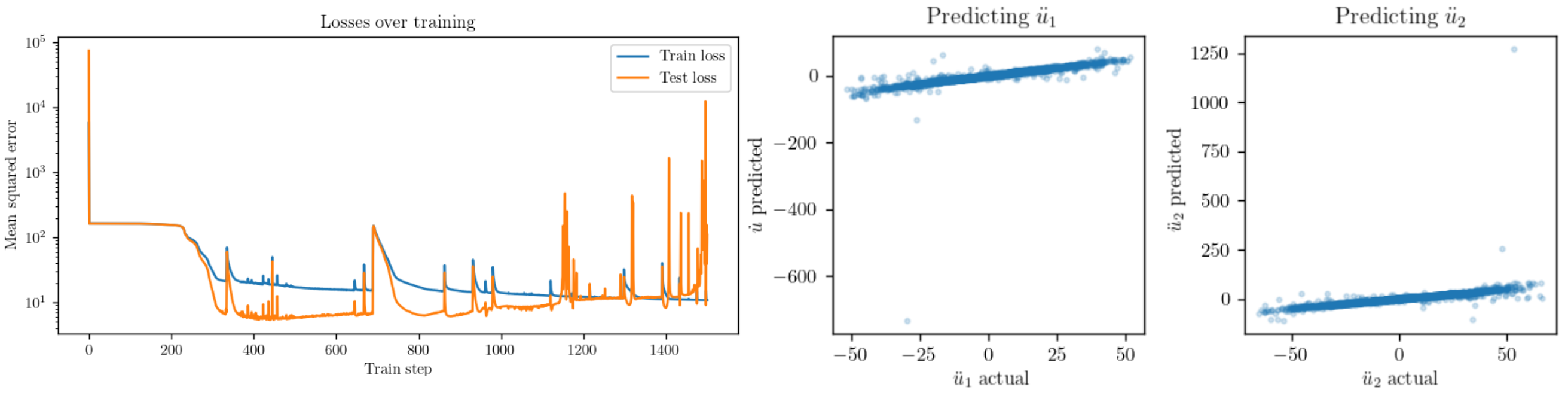}}
\subfigure[The visualization of predicting $\ddot{u}_1$ and $\ddot{u}_2$ over the full coordinate space, where vanilla LNN is trained with noisy data directly.]{
\label{fig:double_pendulum_predition}
\includegraphics[width=1.\textwidth]{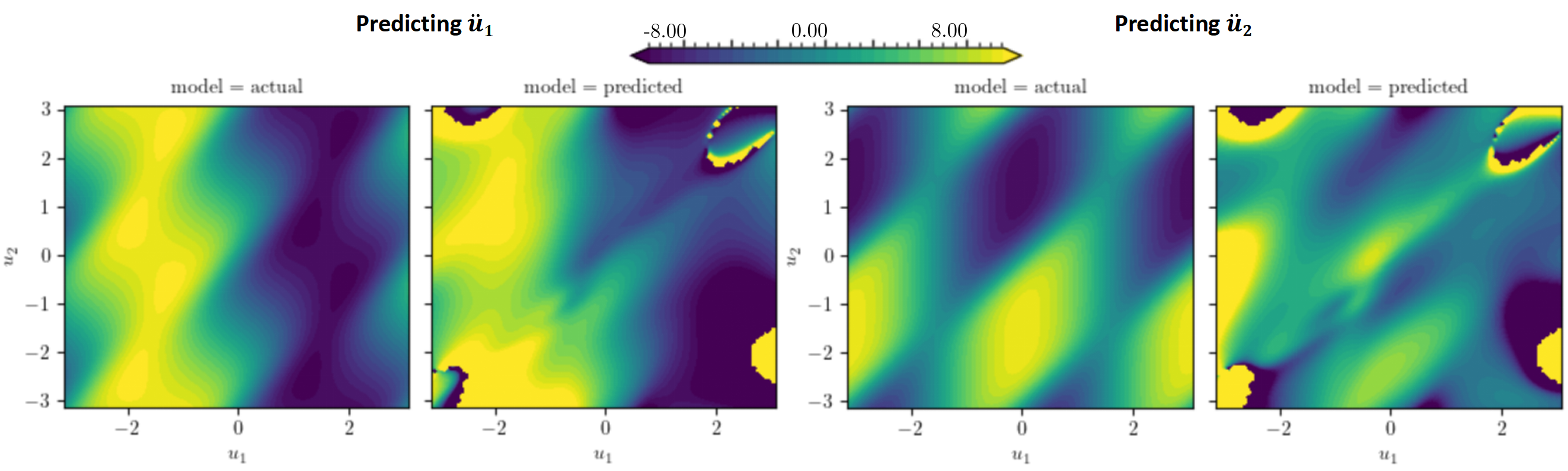}}
\caption{The training and prediction results of vanilla LNN using noisy data directly.}
\label{fig:LNN}
\end{figure}

\begin{figure}[H]
\centering 
\subfigure[The loss curves of training LNN and prediction of $\ddot{u}_1$ and $\ddot{u}_2$ on test dataset using Convex-PIML.]{
\label{fig:double_pendulum_loss_B}
\includegraphics[width=1.\textwidth]{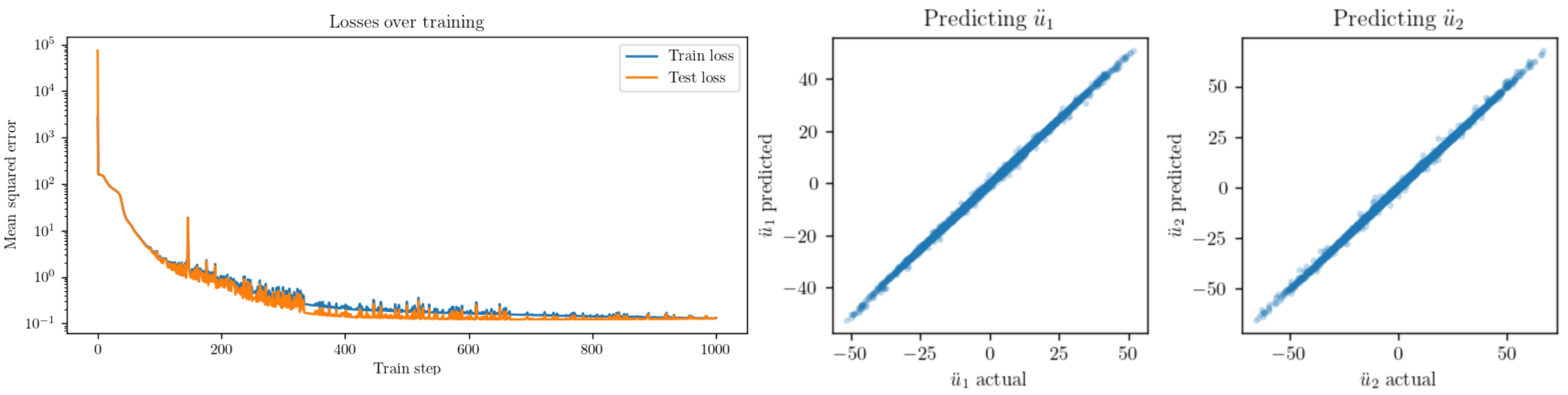}}
\subfigure[The noisy training dataset and approximations of $u_1$ and $u_2$ using B-splines.]{
\label{fig:double_pendulum_approx_B}
\includegraphics[width=1.\textwidth]{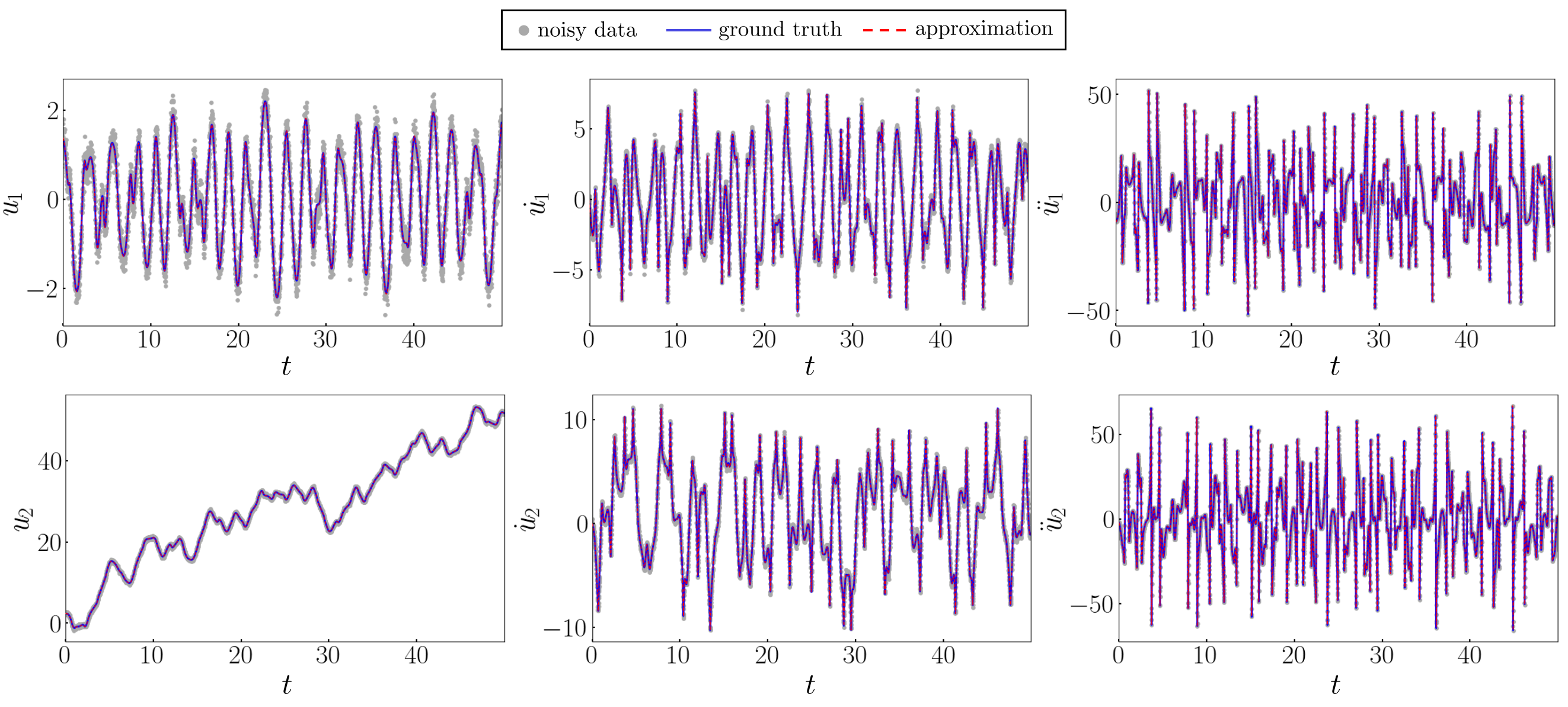}}
\subfigure[The visualization of predicting $\ddot{u}_1$ and $\ddot{u}_2$ over the full coordinate space, where enhanced LNN is trained with Convex-PIML.]{
\label{fig:double_pendulum_predition_B}
\includegraphics[width=1.\textwidth]{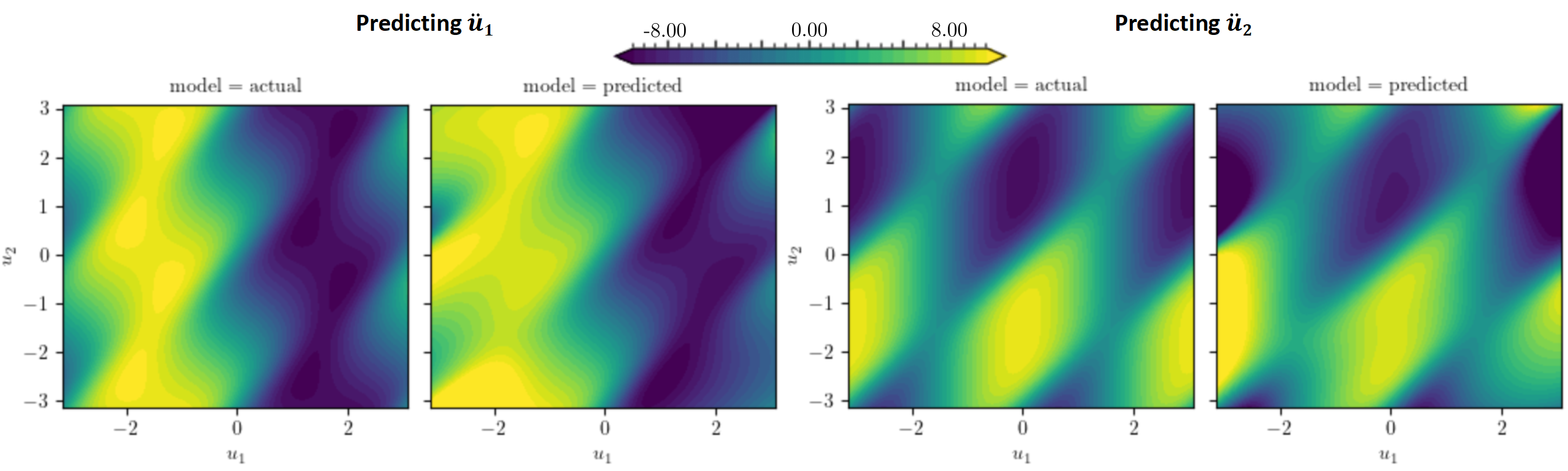}}
\caption{The training and prediction results of enhanced LNN trained with Convex-PIML.}
\label{fig:LNN_B}
\end{figure}

While LNNs have demonstrated promising results using noise-free data in \cite{cranmer2020lagrangian}, we have observed that LNNs are highly sensitive to data quality, particularly to noise. This sensitivity is understandable, given that the Lagrangian $\mathcal{L}(\theta)$ is learned directly from the data, making data quality a critical factor in the performance of LNNs. Since real-world data often contains noise, this sensitivity significantly restricts the practical applicability of LNNs. In this section, we demonstrate how Convex-PIML can mitigate this issue using the example of a double pendulum system, as shown in Fig.\ref{fig:double_pendulum_system}(b).
The noise-free dataset, denoted as $\mathbf{y}=(u_1,\dot{u}_1,\ddot{u}_1,u_2,\dot{u}_2,\ddot{u}_2) \in \mathbb{R}^{10000\times6}$, represents the dynamical states of the double pendulum system over the time interval $t \in [0,100]$ and is obtained through numerical methods. Gaussian noise, sampled from the distribution $\mathcal{N}(0, 0.2^2)$, is introduced into the dataset for the interval $t \in [0,50]$ to create the training dataset, as illustrated by the data points in Fig.\ref{fig:double_pendulum_approx_B}. The noise-free data corresponding to the interval $t \in [50.01, 100]$ is employed as the test dataset. Subsequently, a vanilla LNN is trained using the Adam optimizer on the noisy training dataset. It is observed that the training process exhibited difficulty in convergence and the test loss displayed a tendency to diverge, as depicted in Fig.\ref{fig:double_pendulum_loss}. This observation suggests that the presence of noise severely impedes the training of the LNN, preventing it from accurately learning the correct Lagrangian. Consequently, significant inaccuracies are observed in the predicted data points for $\ddot{u}_1$ and $\ddot{u}_2$ based on the learned Lagrangian. To visualize the predictions for $\ddot{u}_1$ and $\ddot{u}_2$, we begin by randomly sampling 100 values for the angle $u_1$ and 100 values for the angle $u_2$. These sampled angles are then randomly combined, with both $\dot{u}_1$ and $\dot{u}_2$ set to zero, resulting in 10,000 instances of $z = (u_1, u_2, 0, 0)$. These instances are subsequently input into the trained LNN to predict the corresponding values of $\ddot{u}_1$ and $\ddot{u}_2$. The visualization of these predictions over full coordinate space is presented in Fig.\ref{fig:double_pendulum_predition}. When compared to the ground truth, there are evidently unreasonable and non-smooth regions in the predictions. These results indicate that the vanilla LNN is highly sensitive to noise.

Then, we will use the proposed optimization framework to enhance LNN by mitigating this issue. Specifically, we do not input the noisy data into LNN. Instead, we use B-splines to fit the data, and input the smooth approximations into LNN to learn Lagrangian. The loss function is defined as $\mathcal{L}(\beta_1, \beta_2,\theta)=\frac{\lambda_1}{2N}||\mathbf{B}_1\beta_1-u_1||_2^2+\frac{\lambda_2}{2N}||\mathbf{B}_{t1}\beta_1-\dot{u}_1||_2^2+\frac{\lambda_3}{2N}||\mathbf{B}_{tt1}\beta_1-\ddot{u}_1||_2^2+\frac{\lambda_4}{2N}||\mathbf{B}_2\beta_2-u_2||_2^2+\frac{\lambda_5}{2N}||\mathbf{B}_{t2}\beta_2-\dot{u}_2||_2^2+\frac{\lambda_6}{2N}||\mathbf{B}_{tt2}\beta_2-\ddot{u}_2||_2^2+\frac{1}{2N}||\hat{z}(\theta,\beta_1,\beta_2)-z(\beta_1,\beta_2)||^2_2$, where $z(\beta_1,\beta_2)=(\mathbf{B}_1\beta_1,\mathbf{B}_{t1}\beta_1,\mathbf{B}_{2}\beta_2,\mathbf{B}_{t2}\beta_2)$ instead of $(u_1,\dot{u}_1,u_2,\dot{u}_2)$. In this loss function, $\frac{1}{2N}||\hat{z}(\theta,\beta_1,\beta_2)-z(\beta_1,\beta_2)||^2_2$ is the physics loss. Before minimizing the loss function, the B-splines are optimized using the proposed knot optimization. Note that the update of $\theta$ is through the Adam optimizer, since the quadratic approximation is equivalent to gradient descent as mentioned in Methods. The results are depicted in Fig.\ref{fig:LNN_B}. The training and test losses, shown in Fig.\ref{fig:double_pendulum_loss_B}, exhibit smooth convergence. The predictions for $\ddot{u}_1$ and $\ddot{u}_2$ indicate that the enhanced LNN successfully learns the correct Lagrangian. This is evidenced by the fact that the approximations of noisy data using B-splines closely align with the ground truth, as demonstrated in Fig.\ref{fig:double_pendulum_approx_B}. The predictions of $\ddot{u}_1$ and $\ddot{u}_2$ over full coordinate space are also accurate, as illustrated in Fig.\ref{fig:double_pendulum_predition_B}.

The utilization of B-splines for data fitting fundamentally serves the purpose of denoising. A notable advantage of this approach over conventional denoising techniques, such as Fourier transform and wavelet transform, is its ability to ensure that the denoised vector $u$, along with $\dot{u}$ and $\ddot{u}$, are all derived from the same underlying function $u(t)$. This characteristic preserves the intrinsic mathematical consistency across different orders of differentiation, which is often not achievable with traditional methods. Furthermore, Convex-PIML is capable of producing accurate approximations of data after only a few iterations of BSCA, effectively mitigating the divergence issues commonly encountered during the training of LNN. This effectiveness in approximation significantly enhances the robustness of training LNN, making it valuable in handling noisy data. 

\section{Discussion}\label{sec:discussion}
The literature has extensively documented the optimization challenges associated with soft-constrained types of PIML across diverse domains and proposed various remedies. However, no existing approaches fully address all these challenges—such as spectral bias, non-convex optimization, multi-objective optimization, and non-smooth optimization — simultaneously. In this paper, we introduce a distinct framework grounded in convex optimization to effectively resolve PIML optimization problems with high accuracy, efficiency, and stability. Specifically, after imparting a weakly non-convex structure to the loss function via a linear combination of B-splines for data approximation, we then transform the PIML optimization into a series of convex optimization problems using Block Successive Convex Approximation (BSCA).

In this study, we rigorously evaluate the proposed method across a range of scenarios characterized by varying types of physical priors, each designed to achieve specific tasks. These tasks include parameter estimation with known equations, equation discovery, and state prediction governed by conservation law represented by the Euler-Lagrange equation. Our findings demonstrate that all these scenarios derive considerable advantage from the principles of convex optimization. The weakly non-convex nature of the loss function allows for the rapid convergence to a high-quality solution, as the minimization of the convex data fitting loss yields results that are closely aligned with the global optimum. This significantly boosts efficiency, requiring only a limited number of optimization iterations to achieve convergence, as demonstrated in the results. Moreover, the analytical solutions offered by convex optimization can be obtained directly by setting the derivatives to zero, thereby eliminating the need for gradient descent. This results in a jumping update approach for resolving the non-Convex PIML optimization, which consistently avoids local minima and reduces sensitivity to initializations. Consequently, the aforementioned PIML objectives are attained with both precision and robustness. Additionally, the proposed knot optimization method, informed by error estimation, enhances optimization accuracy by addressing the challenge of spectral bias. Given these advantages, we posit that regardless of the parameterization of the physical prior (such as SINDy, symbolic neural networks, etc.), the PIML optimization problem can be effectively solved to achieve a wide array of scientific problems.

Despite these strengths, there are some limitations in the proposed method, primarily concerning the configuration of basis functions. A key limitation is the lack of sufficient localization in the knot refinement strategy. Introducing a new knot often results in the generation of numerous new basis functions within subdomains with small errors, consequently imposing an unnecessary computational burden. Future research should focus on developing more localized knot refinement techniques to prevent excessive refinement of basis functions in regions where satisfactory accuracy has already been achieved. Additionally, the current study employs a rectangular mesh formed by the knots, which is inadequate for addressing problems with irregular boundaries. Incorporating alternative mesh types from the well-established finite element methods could substantially enhance the method's applicability. Moreover, the linear combination of other basis functions, such as Fourier basis functions, could be beneficial in addressing data with periodic characteristics and global structures. Indeed, irrespective of the selection of basis functions, PIML optimization can leverage the advantages inherent in convex optimization. This is because convex optimization provides a robust framework that ensures global optimality and stability in finding solutions, which is particularly beneficial in complex, high-dimensional spaces often encountered in PIML applications. The weakly non-convex nature of the optimization landscape allows for efficient convergence to the optimal solution, thereby enhancing the computational efficiency and reliability of the PIML approach. This characteristic is crucial in ensuring that the optimization process remains tractable and effective, even as the complexity of the underlying physical models increases.
\section*{Acknowledgment}
The research is supported by the Guangdong Provincial Fund - Special Innovation Project (2024KTSCX038) and Research Grants Council of Hong Kong through the Research Impact Fund (R5006-23).


\newpage
\section*{Appendix}
\begin{appendices}
\section{Generating B-spline basis functions using Cox de-Boor Algorithm}
\begin{figure}[H]
\centering
\includegraphics[width=120mm]{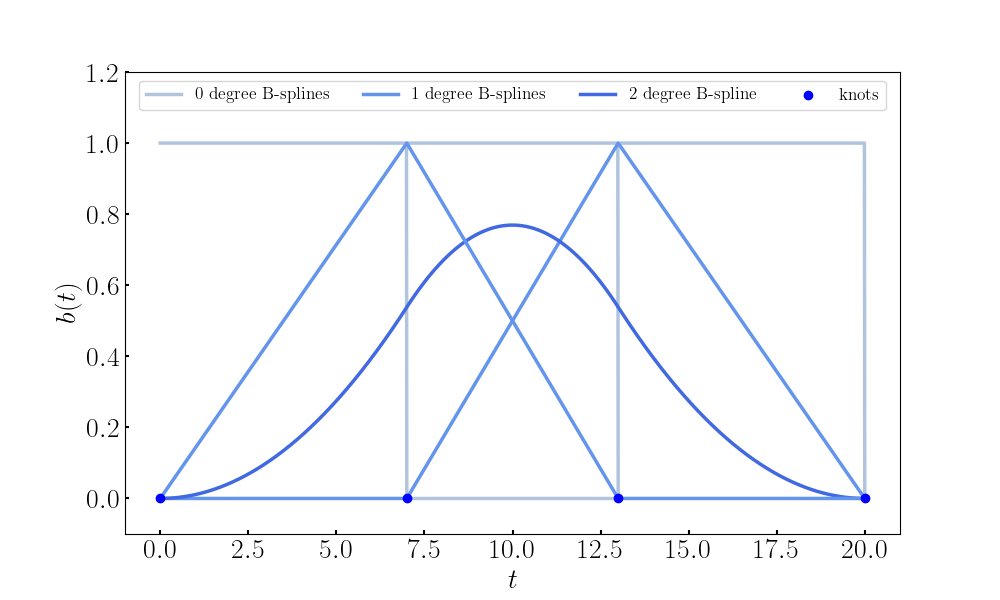}
\caption{The B-splines of degree 0, 1, and 2, generated with uniform knots.}
\label{fig:B-splines}
\end{figure}

The one-dimensional B-splines defined on $[t_0,t_n]$ can be generated by Cox-de Boor algorithm:
\begin{equation}
\label{eq:Cox-de Boor}
\begin{aligned}
   b_{i,0}(t)&=\left\{\begin{aligned}&1\ \ \ \ \text{if}\ t_i\leq t< t_{i+1}\\
&0\ \ \ \ \text{otherwise}\end{aligned}\right.\\
b_{i,g}(t)&=\frac{t-t_i}{t_{i+g}-t_i}b_{i,g-1}(t)+\frac{t_{i+g+1}-t}{t_{i+g+1}-t_{i+1}}b_{i+1,g-1}(t),
\end{aligned}
\end{equation}
where $t_i\;(i=0,1,...,n)$ are the knots; $b_{i,g}(t)$ denotes the $i^{\text{th}}$ B-spline of degree $g$. 
According to Eq.\eqref{eq:Cox-de Boor}, $b_{i,0}(t)$ is local-support: it is nonzero only on interval $[t_i,t_{i+1})$. The $b_{i,g}(t)$ is generated using the $b_{i,g-1}(t)$ and $b_{i+1,g-1}(t)$. According to this recursive rule, $b_{i,g}(t)$ is nonzero only on interval $[t_i,t_{i+g+1})$. Some B-splines of different degrees generated with uniform knots are shown in Fig. \ref{fig:B-splines}, where a B-spline of degree two is generated using two B-splines of degree one, and a B-spline of degree one is generated using two B-splines of degree zero.

\section{A posteriori error estimate of least-squares finite element method}\label{sec:a psteriori}
We leverage the posterior error estimates from R.Verfürth \cite{verfurth1994posteriori,verfurth1996review} to show that the approximation error to the true solution of governing equation is bounded by the physics loss $\mathcal{L}_\text{phy}$.

Let $X,Y$ be two Banach spaces with norm $||\cdot||_X$ and $||\cdot||_Y$. For any element $u\in X$ and any real number $R>0$ set $B(u,R):={v\in X :||u-v||_X<R}$. $\mathcal{L}(X,Y)$ denotes the Banach space of continuous linear maps of $X$ in $Y$ equipped with the operator-norm $||\cdot||_{{\mathcal{L}}(X,Y)}$. $Isom(X,Y)\subset \mathcal{L}(X,Y)$   denotes the open subset of linear homeomorphisms of $X$ onto $Y$. $Y^*:=\mathcal{L}(Y,IR)$ and $<\cdot,\cdot>$ are the dual space of $Y$ and the corresponding duality pairing. Finally, $A^*\in\mathcal{L}(Y^*,X^*)$ denotes the adjoint of a given linear operator $A\in\mathcal{L}(X,Y)$.

Let $F\in C^1(X, Y^*)$ be a given continuously differentiable function. $DF(u)$ is an isomorphism of $X$ onto $Y^*$.The following theorem gives a posteriori error estimates for elements in a neighbourhood of a solution of $F(u)=0$. 

\begin{theorem}[A posteriori error estimate for elements in a neighbourhood of a solution $u$]\label{thm:A posteriori error estimate}
Let $u\in X$ be a regular solution of $F(u)=0$, i.e. $DF(u)\in Isom(X,Y^*)$. Assume that $DF$ is Lipschitz continuous at $u$. Then, the following error estimates hold for all $\hat{u}\in B(u,R)$:
\begin{equation}\label{eq:a posteriori}
   \frac{1}{2}||DF(u)||^{-1}_{\mathcal{L}(X,Y^*)}||F(\hat{u})||_{Y^*} \leq ||u-\hat{u}||_X \leq 2||DF(u)||^{-1}_{\mathcal{L}(Y^*,X)}||F(\hat{u})||_{Y^*}.
\end{equation}
\end{theorem}

We use this a posteriori error estimate to propose a posteriori error estimate of least-squares finite element methods. Let $X,Y$ be two Hilbert spaces. Since Hilbert space is a special type of Banach space, the Theorem \ref{thm:A posteriori error estimate} still holds. Since Hilbert space is self-inverse, $Y^*$ is equivalent to $Y$ and it is also a Hilbert space. Besides, the norms $||\cdot||_{Y^*}$ and $||\cdot||_{X}$ are induced by inner products as Hilbert space is inner product space. In other word, the norms $||\cdot||_{Y^*}$ and $||\cdot||_{X}$ are the 2-norm of a function $||\cdot||_{2}$. Then we have:

\begin{theorem}[A posteriori error estimate of least-squares finite element methods]\label{thm:LSFEM}
Let $u\in X$ be a regular solution of $F(u)=0$, i.e. $DF(u)\in Isom(X,Y^*)$. Assume that $DF$ is Lipschitz continuous at $u$. Then, the following error estimates hold for all $\hat{u}\in H(u,R)$:
\begin{equation}
\frac{1}{2} ||DF(u)||^{-1}_{\mathcal{L}(X,Y)} ||F(\hat{u})||_2 \leq ||u-\hat{u}||_2 \leq 2 ||DF(u)||^{-1}_{\mathcal{L}(Y,X)}||F(\hat{u})||_2.
\end{equation}
\end{theorem}
$||F(\hat{u})||_2$ in Theorem \ref{thm:LSFEM} is the physics loss defined in our study. The theorem implies that the approximation error of true solution $||u-\hat{u}||_2$ is also bounded by the physics loss $||F(\hat{u})||_2$.

\section{The experimental specifics}\label{appendix:The experimental specifics}
\subsection{Parameter estimation with completely known equations}
\subsubsection{K-S equation}
\begin{figure}[H]
\centering 
\subfigure[The optimization results of the B-splines initialization.]{
\label{fig:KS_fitting}
\includegraphics[width=0.7\textwidth]{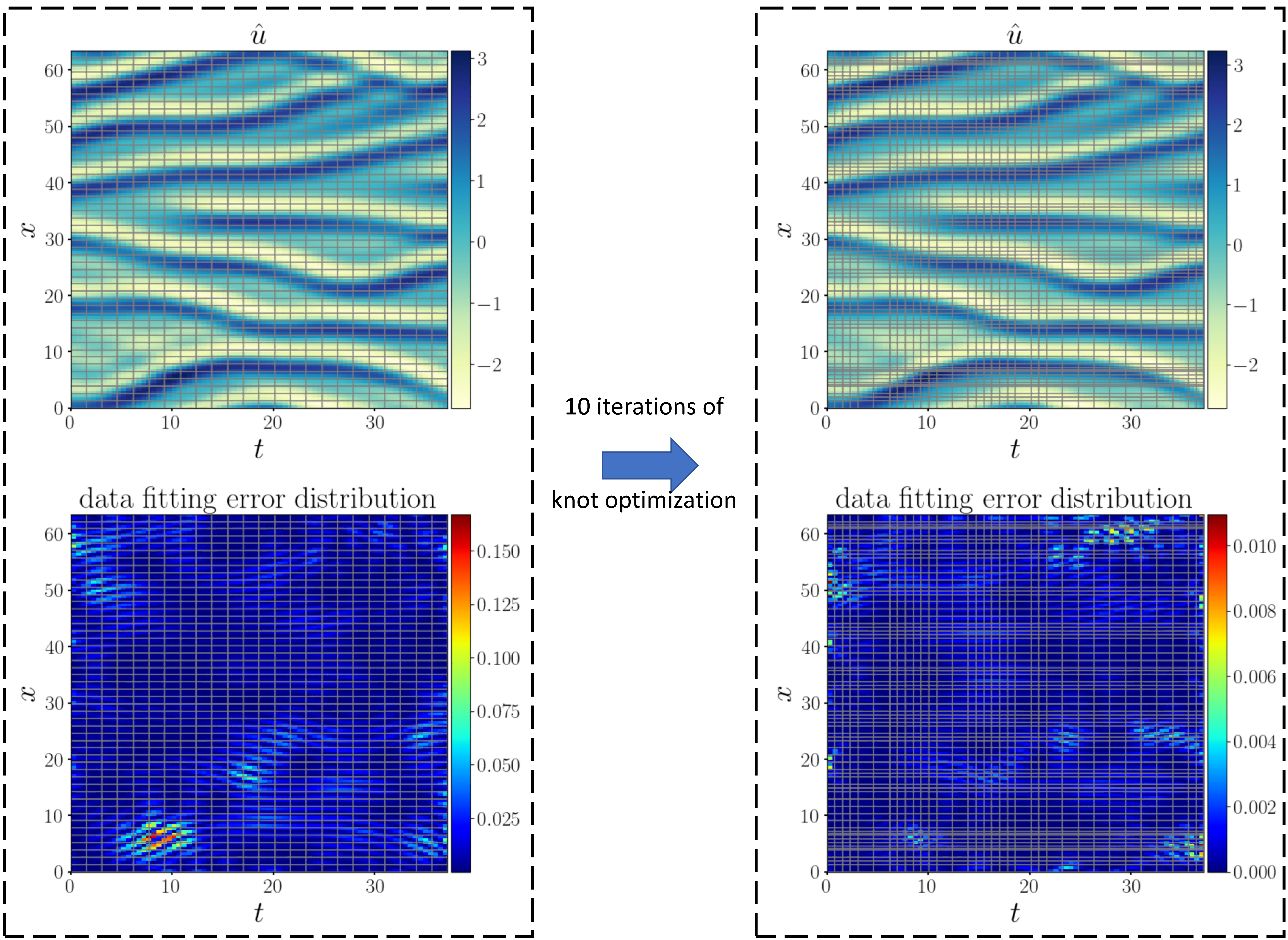}}
\subfigure[The data fitting and physics error distributions before and after the knot optimizations.]{
\label{fig:KS_para_error}
\includegraphics[width=0.7\textwidth]{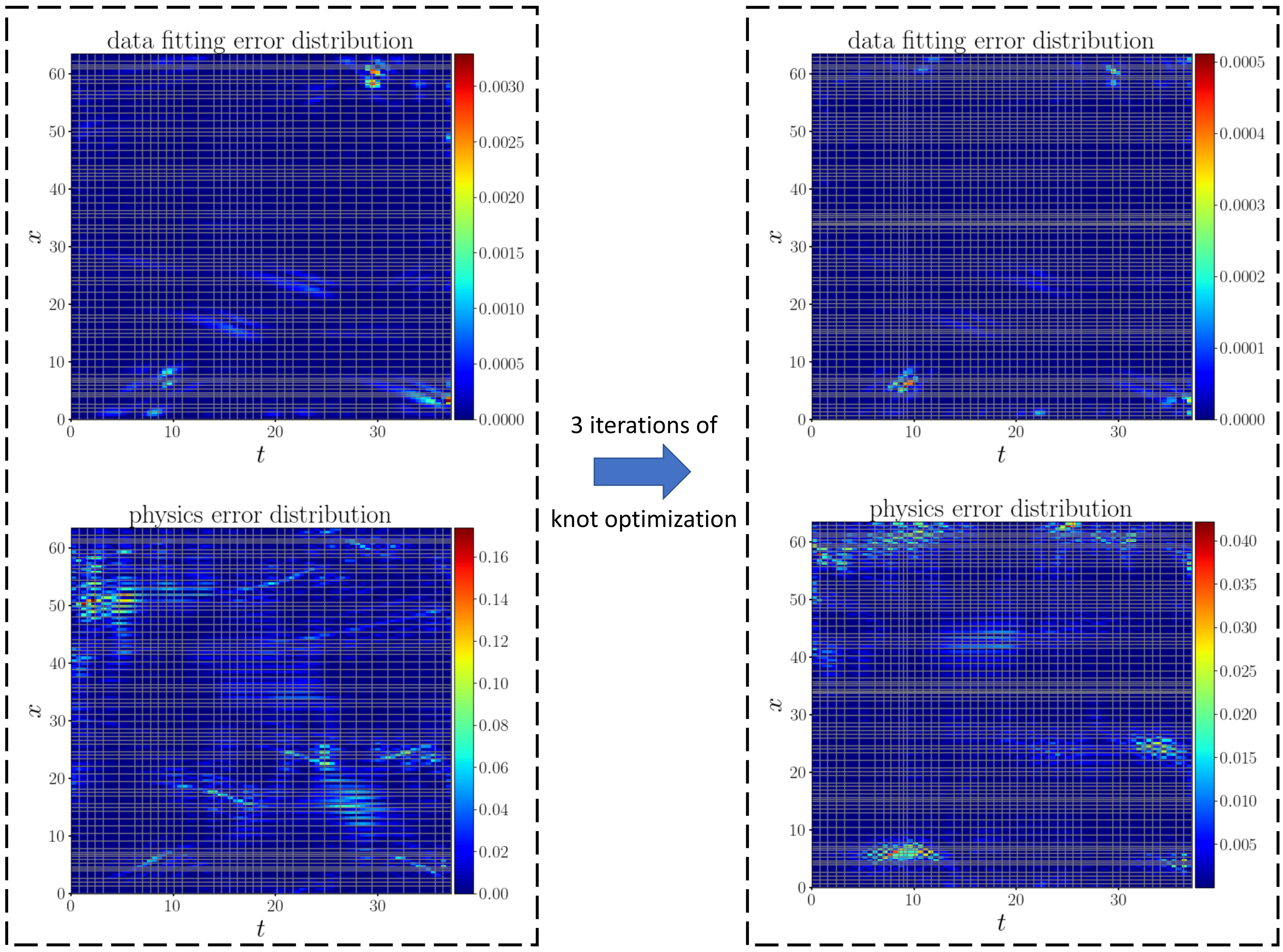}}
\caption{The knot optimization results of B-splines initialization and parameter estimation of K-S equation.}
\label{fig:results_KS_para_est}
\end{figure}

Using the linear combination of B-splines to approximate the solution of K-S equation, the loss function for parameter estimation of K-S equation is given as:
\begin{equation}\label{eq:KS loss function}
\begin{aligned}
    \mathcal{L}(\beta,\theta)&=\frac{\lambda_1}{2N_d}||\mathbf{B}^{d}\beta-\mathbf{y}||^2_2+\frac{\lambda_2}{2N_{ic}}||\mathbf{B}^{ic}\beta-u^{ic}||^2_2+\frac{\lambda_3}{2N_{bc}}||\mathbf{B}^{bc}\beta-u^{bc}||^2_2+\\
    &\frac{1}{2N_c}||\mathbf{B}^{c}_t\beta+\theta_1\mathbf{B}^{c}\beta\circ\mathbf{B}^{c}_x\beta+\theta_2\mathbf{B}^{c}_{xx}\beta+\theta_3\mathbf{B}^{c}_{xxxx}\beta||_2^2,
\end{aligned}
\end{equation}
where $N_d=N_c=9600$ since we let $x^d=x^c$ and $t^d=t^c$ in this example. The trade-off parameters are chosen as $\lambda_1=3,\lambda_2=5,\lambda_3=5$ and fixed at all experiments in this section. Then we use the noise-free data to present the experiment details of parameter estimation of K-S equation. We uniformly distribute $50 \times 25$ knots across the domain $x \in [0, 63.5]$ and $t \in [0, 37]$. Before minimizing this loss function using BSCA, the uniformly distributed knots are first optimized with the knot optimization to improve the data fitting accuracy. Specifically, we solve the pure data fitting problem to obtain data fitting error distribution. Then we can perform knot optimization in each iteration of knot optimization based on the data fitting error distribution. In each iteration of knot optimization, two new knots are inserted in $x$ dimension and one new knot is inserted in $t$ dimension. As shown in Fig. \ref{fig:KS_fitting}, the data fitting error is significantly decreased after the knot optimization. Then we use the optimized knots to generate B-splines to perform parameter estimation of K-S equation.

During the parameter estimation of K-S equation by minimizing the loss function in Eq. \eqref{eq:KS loss function}, three iterations of knot optimization are performed based both data fitting error distribution and physics error distribution. The error distributions before and after knot optimization are presented in Fig. \ref{fig:KS_para_error}, which shows that knot optimization can significantly reduce these two error simultaneously. The above process are repeated five times with different initializations.

\subsubsection{N-S equation}
\begin{figure}[H]
\centering 
\includegraphics[width=0.6\textwidth]{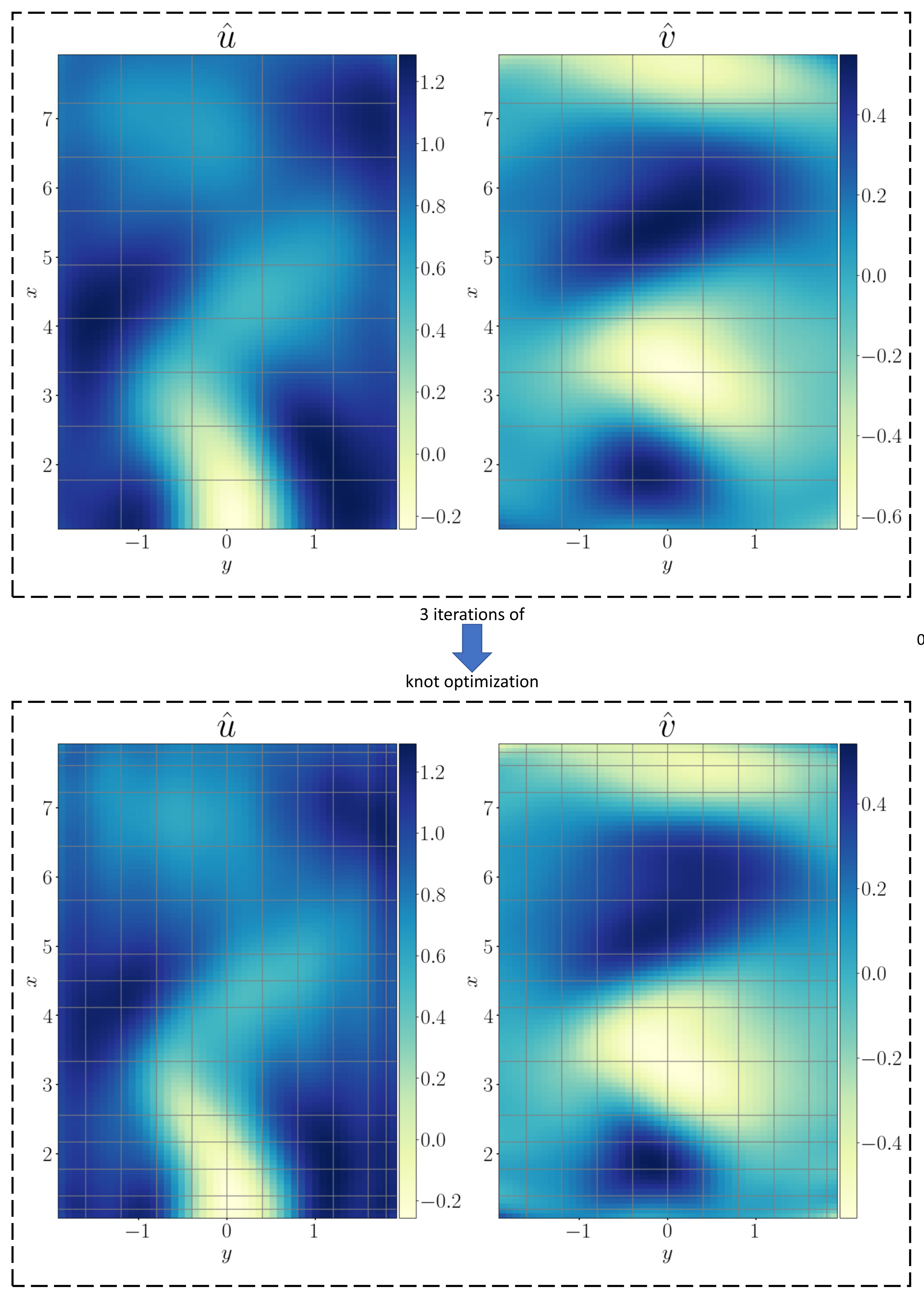}
\caption{The knot optimization results of B-splines initialization of N-S equation.}
\label{fig:NS_fitting}
\end{figure}

Following the derivation of 2D N-S equation expressed in terms of vorticity, the vorticity $\omega$, velocity fields $u$ and $v$ can be represented by stream function $\psi$: $v=-\frac{\partial \psi}{\partial x}$, $u=\frac{\partial \psi}{\partial y}$, $\omega=\frac{\partial v}{\partial x}-\frac{\partial u}{\partial y}=-\frac{\partial^2 \psi}{\partial^2 x}-\frac{\partial^2 \psi}{\partial y^2}$. Therefore, we use B-splines to approximate the stream function $\psi$. Given the data of $u$ and $v$, the loss function for parameter estimation of N-S equation is given as:
\begin{equation}\label{eq:NS loss function}
\begin{aligned}
    \mathcal{L}(\beta,\theta)&=\frac{\lambda_1}{2N^d_u}||\mathbf{B}_y^{d}\beta-u^{d}||^2_2+\frac{\lambda_2}{2N^d_v}||\mathbf{B}_x^{d}\beta+v^{d}||^2_2+\frac{\lambda_3}{2N^{ic}_u}||\mathbf{B}_y^{ic}\beta-u^{ic}||^2_2+\\
    &\frac{\lambda_4}{2N^{ic}_v}||\mathbf{B}_x^{ic}\beta+v^{ic}||^2_2+\frac{\lambda_5}{2N^{bc}_u}||\mathbf{B}_y^{bc}\beta-u^{bc}||^2_2+\frac{\lambda_6}{2N^{bc}_v}||\mathbf{B}_x^{bc}\beta+v^{bc}||^2_2+\\
    &\frac{1}{2N_c}||-(\mathbf{B}^c_{xxt} \beta + \mathbf{B}^c_{yyt} \beta)  - \theta_1 (\mathbf{B}^c_y \beta)\circ (\mathbf{B}^c_{xxx}\beta + \mathbf{B}^c_{xyy}\beta) + \theta_2 (\mathbf{B}^c_x\beta)\circ (\mathbf{B}^c_{xxy}\beta + \mathbf{B}^c_{yyy}\beta) -\\
    &\theta_3(\mathbf{B}^c_{xxxx}\beta + \mathbf{B}^c_{xxyy}\beta) - \theta_4(\mathbf{B}^c_{xxyy}\beta + \mathbf{B}^c_{yyyy}\beta)||_2^2.
\end{aligned}
\end{equation}
Since we divide data into six batches, $N_d=N_c=16666$. The trade-off parameters are chosen as $\lambda_1=100,\lambda_2=1000,\lambda_3=500,\lambda_4=1000,\lambda_5=1000,\lambda_6=500$ and fixed at all experiments in this section. Then we use the noise-free data to present the experiment details of parameter estimation of N-S equation. We uniformly distribute $10 \times 6 \times 4$ knots across the domain defined by $x \in [1.0, 8.0]$, $y \in [-2.0, 2.0]$, and the temporal period $t \in [0, 1.9]$. Before minimizing this loss function using BSCA, the uniformly distributed knots are first optimized with the knot optimization to improve the data fitting accuracy. Specifically, we solve the pure data fitting problem to obtain data fitting error distribution. Then we can perform knot optimization in each iteration of knot optimization based on the data fitting error distribution. In each iteration of knot optimization, a new knot is inserted in $x$ dimension and a new knot is inserted in $y$ dimension based on the data fitting error of $u$; a new knot is inserted in the $x$, $y$, and $t$ dimensions based on the data fitting error of $v$, respectively. The optimized knots are presented in Fig. \ref{fig:NS_fitting}. After knot optimization, the data fitting error of $u$ is decreased from $1.27E-3$ to $9.28E-4$; the data fitting error of $u$ is decreased from $1.01E-3$ to $6.16E-4$. Then we use the optimized knots to generate B-splines to perform parameter estimation of N-S equation. Since there is no multi-scale feature in data of N-S equation, we do not perform knot optimization during the parameter estimation of N-S equation. The above process are repeated five times with different initializations.

\subsection{Equation discovery with partially known equations}
\subsubsection{K-S equation}
A library $\mathbf{\Phi}$ of 35 candidate functions are used to construct the PDE, consisting of $[uu_x, u_{xx}, u_{xxxx} , u_{xxx}, u, u^2, u^3, u^4$, $u^5, u_x, u_{xxxxx},uu_{xx}, uu_{xxx}, uu_{xxxx}, uu_{xxxxx}, u^2u_x$,
$u^2u_{xx}, u^2u_{xxx}, u^2u_{xxxx}, u^2u_{xxxxx}, u^3u_x, u^3u_{xx}, u^3u_{xxx}, u^3u_{xxxx},u^3u_{xxxxx}$,
$u^4u_xu^4u_{xx},u^4u_{xxx}, u^4u_{xxxx}, u^4u_{xxxxx}, u^5u_x, u^5u_{xx}, u^5u_{xxx}, u^5u_{xxxx}, u^5u_{xxxxx}]$. Their initial coefficients are uninformatively chosen to be zeros. The loss function of discovering K-S equation is given as:
\begin{equation}\label{eq:KS loss function for eq disc}
\begin{aligned}
    \mathcal{L}(\beta,\theta)&=\frac{\lambda_1}{2N_d}||\mathbf{B}^{d}\beta-\mathbf{y}||^2_2+\frac{\lambda_2}{2N_{ic}}||\mathbf{B}^{ic}\beta-u^{ic}||^2_2+\frac{\lambda_3}{2N_{bc}}||\mathbf{B}^{bc}\beta-u^{bc}||^2_2+\frac{1}{2N_c}||\mathbf{\Phi}\theta||_2^2+\mu||\theta||_1.
\end{aligned}
\end{equation}

The trade-off parameters $\lambda_1,\lambda_2,\lambda_3$ are the same as the trade-off parameters for parameter estimation of K-S equation, i.e., $\lambda_1=3,\lambda_2=5,\lambda_3=5$. The initialization and the optimization of B-splines are also the same as the trade-off parameters for parameter estimation of K-S equation. The only difference is that there is a $\ell_1$ loss weighted by $\mu=5$. As a result, the updating of $\theta$ is achieved by FISTA since it is a LASSO problem. The diminishing step size is used to update $\beta$, with $\epsilon=0.6$. After determining the correct candidate functions of K-S equation by ST-LASSO, the problem is then transformed into parameter estimation problem.

\subsubsection{N-S equation}
A library $\mathbf{\Phi}$ of 40 candidate functions are used to construct the PDE, consisting of $[u\omega_x, v\omega_y, \omega_{xx}, \omega_{yy}, u, v, uv, u\omega$, $v\omega, u^2, v^2, \omega^2, u\omega_y, u\omega_{xx}, u\omega_{xy}, v\omega_x, v\omega_{xx}, v\omega_{xy},\omega\omega_x, \omega\omega_y, uv\omega_x, uv\omega_y, uv\omega_{xx}, uv\omega_{xy}, u\omega\omega_x, u\omega\omega_y, u\omega\omega_{xx}, u\omega\omega_{xy}$, 
$v\omega\omega_x, v\omega\omega_y, v\omega\omega_{xy}, u^2\omega_x, u^2\omega_y, u^2\omega_{xx}, u^2\omega_{xy}, v^2\omega_x, v^2\omega_y, v^2\omega_{xx}, v^2\omega_{xy}, \omega^2\omega_{xy}]$. Their initial coefficients are uninformatively chosen to be zeros. The loss function of discovering N-S equation is given as:
\begin{equation}\label{eq:NS loss function for eq disc}
\begin{aligned}
    \mathcal{L}(\beta,\theta)&=\frac{\lambda_1}{2N^d_u}||\mathbf{B}_y^{d}\beta-u^{d}||^2_2+\frac{\lambda_2}{2N^d_v}||\mathbf{B}_x^{d}\beta+v^{d}||^2_2+\frac{\lambda_3}{2N^{ic}_u}||\mathbf{B}_y^{ic}\beta-u^{ic}||^2_2+
    \frac{\lambda_4}{2N^{ic}_v}||\mathbf{B}_x^{ic}\beta+v^{ic}||^2_2+\\
    &\frac{\lambda_5}{2N^{bc}_u}||\mathbf{B}_y^{bc}\beta-u^{bc}||^2_2+\frac{\lambda_6}{2N^{bc}_v}||\mathbf{B}_x^{bc}\beta+v^{bc}||^2_2+\frac{1}{2N_c}||\mathbf{\Phi}\theta||_2^2+\mu||\theta||_1.
\end{aligned}
\end{equation}

The trade-off parameters $\lambda_1,\lambda_2,\lambda_3,\lambda_4,\lambda_5,\lambda_6$ are the same as the trade-off parameters for parameter estimation of N-S equation, i.e., $\lambda_1=100,\lambda_2=1000,\lambda_3=500,\lambda_4=1000,\lambda_5=1000,\lambda_6=500$. The initialization and the optimization of B-splines are also the same as the trade-off parameters for parameter estimation of N-S equation. The only difference is that there is a $\ell_1$ loss weighted by $\mu=600$. As a result, the updating of $\theta$ is achieved by FISTA since it is a LASSO problem. The diminishing step size is used to update $\beta$, with $\epsilon=0.9$. After determining the correct candidate functions of K-S equation by ST-LASSO, the problem is then transformed into parameter estimation problem.

\section{The optimal step sizes for parameter estimation using exact line search}\label{appendix:derivation of exact line search}

In this section, we give the optimal step size for the parameter estimation of K-S equation and N-S equation, according to exact line search given as follow: 
\begin{equation}
\begin{aligned}
        \gamma^{(k)}&=\mathop{\arg \min}\limits_{0\leq\gamma\leq1}\ \gamma(g(\tilde{\beta}^{(k)})-g(\beta^{(k-1)}))+h(\beta^{(k-1)}+\gamma(\tilde{\beta}^{(k)}-\beta^{(k-1)}),\theta^{(k-1)}).
\end{aligned}
\end{equation}
\subsection{K-S equation}\label{apd:optimal step size for K-S equation}
The loss function is given in Eq. \eqref{eq:KS loss function}, where $g(\beta)=\frac{\lambda_1}{2N_d}||\mathbf{B}^{d}\beta-\mathbf{y}||^2_2+\frac{\lambda_2}{2N_{ic}}||\mathbf{B}^{ic}\beta-u^{ic}||^2_2+\frac{\lambda_3}{2N_{bc}}||\mathbf{B}^{bc}\beta-u^{bc}||^2_2$, $h(\beta,\theta)=\frac{1}{2N_c}||\mathbf{B}_t\beta+\theta_1\mathbf{B}\beta\circ\mathbf{B}_x\beta+\theta_2\mathbf{B}_{xx}\beta+\theta_3\mathbf{B}_{xxxx}\beta||_2^2$. Let $\Delta=\tilde{\beta}^{(k)}-\beta^{(k-1)}$, $\text{A}_0=\frac{\lambda_1}{2N_d}(||\mathbf{B}\tilde{\beta}^{(k)}-\mathbf{y}||_2^2-||\mathbf{B}\beta^{(k-1)}-\mathbf{y}||_2^2)+\frac{\lambda_2}{2N_{ic}}(||\mathbf{B}^{ic}\tilde{\beta}^{(k)}-u^{ic}||_2^2-||\mathbf{B}^{ic}\beta^{(k-1)}-u^{ic}||_2^2)+\frac{\lambda_3}{2N_{bc}}(||\mathbf{B}^{bc}\tilde{\beta}^{(k)}-u^{bc}||_2^2-||\mathbf{B}\beta^{(k-1)}-u^{bc}||_2^2),\ \mathbf{A}_{1}=\theta^{(k-1)}_{1}\mathbf{B}\Delta\mathbf{\odot}\mathbf{B}_{x}\Delta,\ \mathbf{A}_{2}=\mathbf{B}_{t}\Delta+\theta^{(k-1)}_{1}\mathbf{B}\beta^{(k-1)}\odot\mathbf{B}_{x}\Delta+\theta^{(k-1)}_{1}\mathbf{B}\Delta\mathbf{\odot}\mathbf{B}_{x}\beta^{(k-1)}+\theta^{(k-1)}_{2}\mathbf{B}_{xx}\Delta+\theta^{(k-1)}_{3}\mathbf{B}_{xxxx}\Delta,\ \mathbf{A}_{3}=\mathbf{B}_{t}\beta^{(k-1)}+\theta^{(k-1)}_{1}\mathbf{B}\beta^{(k-1)}\odot\mathbf{B}_{x}\beta^{(k-1)}+\theta^{(k-1)}_{2}\mathbf{B}_{xx}\beta^{(k-1)}+\theta^{(k-1)}_{3}\mathbf{B}_{xxxx}\beta^{(k-1)}$, we have:
\begin{equation}
    \begin{aligned}
\gamma^{(k)}&=\arg\min_{0\leq\gamma\leq1} \gamma \text{A}_0+\frac{1}{2N_c}(\gamma^2\mathbf{A}_1^T+\gamma\mathbf{A}_2^T+\mathbf{A}_3^\mathrm{T})(\gamma^2\mathbf{A}_1+\gamma\mathbf{A}_2+\mathbf{A}_3)\\
=&\mathop{\arg} \min \limits_{0\leq\gamma\leq1}\ \frac{1}{2N_c}\mathbf{A}_1^T\mathbf{A}_1\gamma^4+\frac{1}{2N_c}(\mathbf{A}_1^T\mathbf{A}_2+\mathbf{A}_2^T\mathbf{A}_1)\gamma^3+\\
&\frac{1}{2N_c}(\mathbf{A}_1^T\mathbf{A}_3+\mathbf{A}_2^T\mathbf{A}_2+\mathbf{A}_3^T\mathbf{A}_1)\gamma^2+(\frac{1}{2N_c}\mathbf{A}_2^T\mathbf{A}_3+\frac{1}{2N_c}\mathbf{A}_3^T\mathbf{A}_2+\text{A}_0)\gamma+\frac{1}{2N_c}\mathbf{A}_3^T\mathbf{A}_3
    \end{aligned}
\end{equation}

Then we can obtain the optimal step size for the $k^{th}$ iteration by setting the derivative to zero, which is to solve a cubic equation:
\begin{equation}
\frac{2}{N_c}\mathbf{A}_1^T\mathbf{A}_1\gamma^3+\frac{3}{2N_c}(\mathbf{A}_1^T\mathbf{A}_2+\mathbf{A}_2^T\mathbf{A}_1)\gamma^2+\frac{1}{2N_c}(\mathbf{A}_1^T\mathbf{A}_3+\mathbf{A}_2^T\mathbf{A}_2+\mathbf{A}_3^T\mathbf{A}_1)\gamma+(\frac{1}{2N_c}\mathbf{A}_2^T\mathbf{A}_3+\frac{1}{2N_c}\mathbf{A}_3^T\mathbf{A}_2+\text{A}_0)=0
\end{equation}

\subsection{N-S equation}
The loss function is given in Eq. \eqref{eq:NS loss function}, where $g(\beta)=\frac{\lambda_1}{2N^d_u}||\mathbf{B}_y^{d}\beta-u^{d}||^2_2+\frac{\lambda_2}{2N^d_v}||\mathbf{B}_x^{d}\beta+v^{d}||^2_2+\frac{\lambda_3}{2N^{ic}_u}||\mathbf{B}_y^{ic}\beta-u^{ic}||^2_2+\frac{\lambda_4}{2N^{ic}_v}||\mathbf{B}_x^{ic}\beta+v^{ic}||^2_2+\frac{\lambda_5}{2N^{bc}_u}||\mathbf{B}_y^{bc}\beta-u^{bc}||^2_2+\frac{\lambda_6}{2N^{bc}_v}||\mathbf{B}_x^{bc}\beta+v^{bc}||^2_2$, $h(\beta,\theta)=\frac{1}{2N_c}||-(\mathbf{B}_{xxt} \beta + \mathbf{B}_{yyt} \beta)  - \theta_1 (\mathbf{B}_y \beta)(\mathbf{B}_{xxx}\beta + B_{xyy}\beta) + \theta_2 (\mathbf{B}_x\beta)(\mathbf{B}_{xxy}\beta + \mathbf{B}_{yyy}\beta) -\theta_3(B_{xxxx}\beta + B_{xxyy}\beta) - \theta_4(B_{xxyy}\beta + B_{yyyy}\beta)||_2^2$. 

Let $\Delta=\tilde{\beta}^{(k)}-\beta^{(k-1)}$, $\text{A}_0=\frac{\lambda_1}{2N^d_u}(||\mathbf{B}_y^{d}\tilde{\beta}^{(k)}-u^{d}||^2_2-||\mathbf{B}_y^{d}\beta^{(k-1)}-u^{d}||^2_2)+\frac{\lambda_2}{2N^d_v}(||\mathbf{B}_x^{d}\tilde{\beta}^{(k)}+v^{d}||^2_2-||\mathbf{B}_x^{d}\beta^{(k-1)}+v^{d}||^2_2)+\frac{\lambda_3}{2N^{ic}_u}(||\mathbf{B}_y^{ic}\tilde{\beta}^{(k)}-u^{ic}||^2_2-||\mathbf{B}_y^{ic}\beta^{(k-1)}-u^{ic}||^2_2)+\frac{\lambda_4}{2N^{ic}_v}(||\mathbf{B}_x^{ic}\tilde{\beta}^{(k)}+v^{ic}||^2_2-||\mathbf{B}_x^{ic}\beta^{(k-1)}+v^{ic}||^2_2)+\frac{\lambda_5}{2N^{bc}_u}(||\mathbf{B}_y^{bc}\tilde{\beta}^{(k)}-u^{bc}||^2_2-||\mathbf{B}_y^{bc}\beta^{(k-1)}-u^{bc}||^2_2)+\frac{\lambda_6}{2N^{bc}_v}(||\mathbf{B}_x^{bc}\tilde{\beta}^{(k)}+v^{bc}||^2_2-||\mathbf{B}_x^{bc}\beta^{k-1}+v^{bc}||^2_2)$, $\mathbf{A}_{1}=\theta^{(k-1)}_2\mathbf{B}_x\Delta\odot(\mathbf{B}_{xxy} + \mathbf{B}_{yyy})\Delta - \theta^{(k-1)}_1\mathbf{B}_y\Delta\odot(\mathbf{B}_{xxx} + \mathbf{B}_{xyy}) \Delta,\ \mathbf{A}_{2}=\theta^{(k-1)}_2\mathbf{B}_x\beta \odot (\mathbf{B}_{xxy} + \mathbf{B}_{yyy}) \Delta + \theta^{(k-1)}_2\mathbf{B}_x \Delta \odot (\mathbf{B}_{xxy} + \mathbf{B}_{yyy})\beta + \theta^{(k-1)}_3 (\mathbf{B}_{xxxx} + \mathbf{B}_{xxyy})\Delta + \theta^{(k-1)}_4(\mathbf{B}_{xxyy} + \mathbf{B}_{yyyy}) \Delta -\theta^{(k-1)}_1\mathbf{B}_y \beta \odot (\mathbf{B}_{xxx} + \mathbf{B}_{xyy}) \Delta - \theta^{(k-1)}_1\mathbf{B}_y \Delta \odot (\mathbf{B}_{xxx} + \mathbf{B}_{xyy})\beta - (\mathbf{B}_{xxt} + \mathbf{B}_{yyt}) \Delta,\mathbf{A}_{3}=\theta^{(k-1)}_2\mathbf{B}_x\beta\odot (\mathbf{B}_{xxy} + \mathbf{B}_{yyy})\beta + \theta^{(k-1)}_3 (\mathbf{B}_{xxxx} + \mathbf{B}_{xxyy})\beta + \theta^{(k-1)}_4 (\mathbf{B}_{xxyy} +
\mathbf{B}_{yyyy})\beta- \theta^{(k-1)}_1\mathbf{B}_y\beta \odot (\mathbf{B}_{xxx} + \mathbf{B}_{xyy}) \beta - (\mathbf{B}_{xxt} + \mathbf{B}_{yyt})\beta$, we have:
\begin{equation}
    \begin{aligned}
\gamma^{(k)}&=\arg\min_{0\leq\gamma\leq1} \gamma \text{A}_0+\frac{1}{2N_c}(\gamma^2\mathbf{A}_1^T+\gamma\mathbf{A}_2^T+\mathbf{A}_3^\mathrm{T})(\gamma^2\mathbf{A}_1+\gamma\mathbf{A}_2+\mathbf{A}_3)\\
=&\mathop{\arg} \min \limits_{0\leq\gamma\leq1}\ \frac{1}{2N_c}\mathbf{A}_1^T\mathbf{A}_1\gamma^4+\frac{1}{2N_c}(\mathbf{A}_1^T\mathbf{A}_2+\mathbf{A}_2^T\mathbf{A}_1)\gamma^3+\\
&\frac{1}{2N_c}(\mathbf{A}_1^T\mathbf{A}_3+\mathbf{A}_2^T\mathbf{A}_2+\mathbf{A}_3^T\mathbf{A}_1)\gamma^2+(\frac{1}{2N_c}\mathbf{A}_2^T\mathbf{A}_3+\frac{1}{2N_c}\mathbf{A}_3^T\mathbf{A}_2+\text{A}_0)\gamma+\frac{1}{2N_c}\mathbf{A}_3^T\mathbf{A}_3
    \end{aligned}
\end{equation}
Then we can obtain the optimal step size for the $k^{th}$ iteration by setting the derivative to zero, which is to solve a cubic equation as well:
\begin{equation}
\frac{2}{N_c}\mathbf{A}_1^T\mathbf{A}_1\gamma^3+\frac{3}{2N_c}(\mathbf{A}_1^T\mathbf{A}_2+\mathbf{A}_2^T\mathbf{A}_1)\gamma^2+\frac{1}{2N_c}(\mathbf{A}_1^T\mathbf{A}_3+\mathbf{A}_2^T\mathbf{A}_2+\mathbf{A}_3^T\mathbf{A}_1)\gamma+(\frac{1}{2N_c}\mathbf{A}_2^T\mathbf{A}_3+\frac{1}{2N_c}\mathbf{A}_3^T\mathbf{A}_2+\text{A}_0)=0
\end{equation}

\end{appendices}

\printcredits

\bibliographystyle{cas-model2-names}
\bibliography{cas-refs}


\end{document}